\documentclass[preprint,10pt]{elsarticle}

\usepackage[a4paper,margin=1in]{geometry}
\usepackage{placeins}
\usepackage{booktabs}

\usepackage{amssymb}
\usepackage{hyperref}
\usepackage{amsmath}
\usepackage{amsthm}
\usepackage{bbm}

\usepackage{ragged2e}
\usepackage{array}

\usepackage{caption} 
\usepackage{threeparttable}

\newtheoremstyle{myremark}
  {3pt}
  {3pt}
  {}
  {}
  {\bfseries}
  {.}
  { }
  {}

\theoremstyle{myremark}

\DeclareMathOperator*{\argmin}{arg\,min}

\newcommand\at[2]{\left.#1\right|_{#2}}

\journal{}

\begin{document}

\begin{frontmatter}



\title{A physics-informed data-driven framework for modeling hyperelastic materials with progressive damage and failure}


\author{Kshitiz Upadhyay\corref{cor1}}
\cortext[cor1]{Corresponding author. Email: kshitizu@umn.edu}
\address{Department of Aerospace Engineering and Mechanics, University of Minnesota, Minneapolis, MN 55455, USA}
\begin{abstract}
This work presents a two-stage physics-informed, data-driven constitutive modeling framework for hyperelastic soft materials undergoing progressive damage and failure. The framework is grounded in the concept of hyperelasticity with energy limiters and employs Gaussian Process Regression (GPR) to separately learn the intact (undamaged) elastic response and damage evolution directly from data. In Stage I, GPR models learn the intact hyperelastic response through volumetric and isochoric response functions (or only the isochoric response under incompressibility), ensuring energetic consistency of the intact response and satisfaction of fundamental principles such as material frame indifference and balance of angular momentum. In Stage II, damage is modeled via a separate GPR model that learns the mapping between the intact strain energy density predicted by Stage I models and a stress-reduction factor governing damage and failure, with monotonicity, non-negativity, and complete-failure constraints enforced through penalty-based optimization to ensure thermodynamic admissibility. Validation on synthetic datasets, including benchmarking against analytical constitutive models and competing data-driven approaches, demonstrates high in-distribution accuracy under uniaxial tension and robust generalization from limited training data to compression and shear modes not used during training. Application to experimental brain tissue data demonstrates the practical applicability of the framework and enables inference of damage evolution and critical failure energy. Overall, the proposed framework combines the physical consistency, interpretability, and generalizability of analytical models with the flexibility, predictive accuracy, and automation of machine learning, offering a powerful approach for modeling failure in soft materials under limited experimental data.
\end{abstract}

\begin{keyword}
Data-driven constitutive models \sep Hyperelasticity \sep Damage and failure \sep Physics-informed machine learning  \sep Gaussian process regression \sep Soft materials \sep Brain tissue mechanics

\end{keyword}

\end{frontmatter}


\section{Introduction}
\label{sec:sample1}

Hyperelastic materials, such as elastomers, hydrogels, rubbers, and foams, are widely utilized in diverse applications ranging from biomedical devices and soft robotics to automotive components and protective equipment. These soft materials typically exhibit nonlinear stress–strain responses and undergo large deformations, complicating the modeling of their mechanical behavior \cite{Puglisi2016,volokh2016mechanics}. Moreover, soft materials often experience progressive damage leading to eventual failure under significant deformation, an aspect not captured by conventional hyperelastic models \cite{Avril2017,Xiang2020,Volokh2010}. Accurately representing both hyperelastic and damage behaviors in constitutive models is thus crucial for reliable predictive simulations of real-world mechanical responses \cite{Volokh2010,Konale2025,Chittajallu2022,Chandrashekar2025}.

Traditional constitutive models, rooted in continuum thermodynamics, have effectively described the mechanical responses of various hyperelastic materials. Prominent examples include neo-Hookean \cite{Treloar1943,Rivlin1948a}, Mooney--Rivlin \cite{Mooney1940,Rivlin1948b}, Ogden \cite{Ogden1972}, and Gent models \cite{Gent1996}, appreciated for their physical interpretability and mathematical simplicity. These traditional models inherently satisfy fundamental physical constraints such as objectivity (i.e.,  material frame indifference), material symmetry, and thermodynamic consistency \cite{Holzapfel2000,Upadhyay2019}. However, their fixed mathematical forms constrain their flexibility, rendering them insufficient for accurately capturing highly complex and varied responses \cite{Destrade2017}. Additionally, these models often require expert-driven selection tailored to specific stress–strain data, creating a significant bottleneck for automated and high-throughput material characterization and discovery \cite{Upadhyay2020,Beda2014,Wex2015,Khaniki2023}. Consequently, there is an urgent need for flexible data-driven constitutive modeling methods that can autonomously discover relationships directly from data with minimal expert intervention.

Data-driven constitutive modeling has emerged as a powerful alternative, overcoming many limitations inherent in traditional approaches \cite{Fuhg2025,Kumar2022}. These methods autonomously identify flexible constitutive models directly from experimental or numerical data, significantly reducing dependence on expert input \cite{Chen2021,Tang2020,Tang2019,Lu2020,Asad2022,Ostadrahimi2026b}. Moreover, data-driven approaches can effectively generalize across diverse material responses, providing universally applicable solutions ideal for automated characterization pipelines \cite{Peirlinck2024,He2021,Linka2021,Kalina2022,Masi2023}. Concurrently, advancements in experimental techniques, including digital image correlation (DIC), digital volume correlation (DVC), and magnetic resonance elastography (MRE), have significantly enriched the available strain data, further enhancing the efficacy of data-driven methods \cite{Flaschel2021,Flaschel2022,Hu2025,Jeong2024,Eggersmann2019,Kirchdoerfer2016,Yang2021}. These techniques facilitate higher accuracy, improved generalizability, and reduced reliance on restrictive assumptions.

Data-driven constitutive modeling approaches can broadly be classified into purely data-driven (black-box) methods and physics-informed data-driven frameworks (for details, refer to a recent review by Fuhg et al. \cite{Fuhg2025}). Purely data-driven methods, such as standard artificial neural networks (ANNs), provide considerable flexibility and predictive accuracy within training datasets. However, their practical engineering utility is often limited by poor generalizability and insufficient compliance with fundamental physical constraints. In contrast, physics-informed data-driven models explicitly incorporate essential physical principles, ensuring adherence to critical constraints such as objectivity, material symmetry, polyconvexity, and thermodynamic consistency.

Recent research on mechanics of soft materials has highlighted various physics-informed machine learning (ML) approaches, including physics-informed neural networks (PINNs), Gaussian Process Regression (GPR)-based surrogate models, and Input Convex Neural Networks (ICNNs) \cite{Klein2022,Thakolkaran2025,Wang2021,Fuhg2022,Fuhg2022b,Frankel2020,Liu2020,Konale2026,Ostadrahimi2026}. For instance, studies by Frankel et al. \cite{Frankel2020} and Fuhg et al. \cite{Fuhg2022b} represented the total stress in a generalized hyperelastic constitutive framework as a weighted sum of components from an irreducible tensor basis and employed GPR to map strain invariants to the corresponding coefficients of these basis components—referred to as response functions. This simple mapping between invariants and response functions inherently enforces key physical constraints on the predicted mechanical responses, including material frame indifference, material symmetry, thermodynamic consistency, and the local balance of angular momentum. Furthermore, unlike other regression methods (e.g., ANNs), GPR offers the benefit of enforcing a stress-free reference configuration through its exact inference property \cite{Fuhg2022}. Extending this line of work, Upadhyay et al. \cite{Upadhyay2024} recently applied a constrained GPR framework to visco-hyperelastic materials, explicitly integrating entropy inequality constraints into GPR hyperparameter optimization. This work highlighted several additional benefits of GPR-based data-driven models: compatibility with limited training data, generalizability across a wide range of deformation modes, robustness to noise, and enhanced interpretability owing to their rigorous statistical foundation, which also facilitates seamless integration with uncertainty quantification pipelines.

Despite substantial advancements, a critical gap remains in capturing progressive damage and failure behaviors within hyperelastic modeling frameworks, both in traditional and data-driven modeling. Conventional hyperelastic theories inherently fail to represent material failure adequately, as their strain energy density functions typically grow without bound, thereby limiting realistic predictions of material performance. Addressing this limitation, Volokh introduced the concept of energy limiters, which embeds a saturation limit into the strain energy density function, inherently enabling the representation of material failure at large deformations \cite{Volokh2007}. This energy-limiting concept has effectively modeled various phenomena, including rubber failure under diverse loading conditions, crack initiation and propagation, cavitation instabilities, and membrane rupture \cite{Volokh2017,Abu-Qbeitah2023,Faye2019,Lev2019}. However, as described earlier, such classical analytical approaches heavily depend on carefully chosen functional forms and expert assumptions, constraining their flexibility and generalization capabilities.

Efforts to integrate data-driven methods with damage mechanics have mainly focused on neural-network-based architectures. For example, Tac et al. \cite{Tac2024} employed neural ordinary differential equations (NODEs) to model hyperelastic materials and extended this approach to capture damage in a thermodynamically consistent manner by introducing monotonic yield functions. Their approach demonstrated the ability to rediscover classical damage evolution laws from synthetic data and to describe soft-tissue responses under large deformations. Beyond NODE-based approaches, several studies have explored neural networks to learn latent or anisotropic damage representations. Gaetano et al. \cite{Gaetano2025} used deep neural networks to infer a fourth-order anisotropic damage surface tensor in fiber-reinforced composites from RVE simulations, linking macroscopic damage evolution to applied macrostrains and secant stiffness degradation. Liu et al. \cite{Liu2025} trained ANNs to construct damage-form yield functions for plate and shell fracture using invariants of the effective active stress. Rojas et al. \cite{Rojas2023} investigated neural architectures—including PINNs—to infer parameters in phase-field damage PDEs. Parallel to these neural-network-based constitutive approaches, other recent studies have begun to explore alternative data-driven tools: Ostadrahimi et al. \cite{Ostadrahimi2026} combined GPR and recurrent neural networks to characterize Mullins-type discontinuous softening in elastomers, while Abdusalamov et al. \cite{Abdusalamov2024} used symbolic regression to model this stress-softening phenomenon. Yet another class of studies has employed ML primarily as a high-dimensional black-box representation or as a computational surrogate for fracture and damage simulations \cite{Ani2026,Konale2026}. These contributions collectively underscore a broader research trend toward learning damage evolution laws directly from data; however, most existing frameworks either rely on high-capacity neural architectures that may limit interpretability and generalizability across varied deformation scenarios, focus on accelerating fracture simulations rather than learning local constitutive laws, or target discontinuous softening phenomena (e.g., Mullins effect, hysteresis, residual strain) rather than continuous stiffness degradation along the loading path leading to mechanical failure.

Motivated by these gaps and opportunities, the present work introduces a novel physics-informed data-driven framework for modeling continuous, mechanically driven damage of hyperelastic materials leading to failure, characterized by progressive stiffness reduction along the primary loading path and culminating in material rupture. Building upon Volokh's robust physical foundations of hyperelasticity with energy limiters \cite{Volokh2007,Volokh2013}, our approach employs a generalized functional form that expresses stress through irreducible tensor basis components weighted by hyperelastic response functions and modulated by a stress-reduction factor dependent on intact (i.e., undamaged) strain energy density. In contrast to prior neural-network-based constitutive approaches, the present work adopts a GPR-based two-stage supervised learning strategy, initially capturing undamaged hyperelastic response functions, followed by learning damage response through a strain-energy-density-dependent stress-reduction factor. Crucially, the second stage enforces physically motivated constraints, including monotonic, non-negative damage accumulation and complete failure at large strains.

The proposed model is comprehensively validated using numerically generated stress-strain datasets. Predictions are benchmarked against traditional phenomenological models, black-box ML (i.e., purely data-driven) models, and direct one-step GPR mappings across diverse deformation modes, including uniaxial and shear responses extending beyond the training regime. As part of this evaluation, we also applied our framework to experimentally observed stress–strain data of brain tissue exhibiting progressive damage, further demonstrating the model’s applicability to complex biological materials. Results indicate substantial improvements in accuracy, predictive robustness, and generalizability.

The remainder of this paper is structured as follows: Section \ref{sec:Theoretical_Framework} outlines the theoretical foundations, including the generalized energy-limited hyperelastic model and its irreducible tensor basis decomposition. Section \ref{sec:proposed_modeling_framework} details the proposed data-driven constitutive modeling framework and our two-stage constrained training approach. Section \ref{sec:Model_eval_method} describes model evaluation methodology and the numerical and experimental datasets used to that end. Section \ref{sec:results_and_discussion} presents and analyzes model performance and evaluation results, and Section \ref{sec:summary_and_conclusion} summarizes key findings and suggests future research directions.

\section{Theoretical Foundations}
\label{sec:Theoretical_Framework}

\subsection{Energy Limiters Approach to Modeling Failure in Hyperelastic Materials}

Traditional hyperelastic material models employ strain energy density functions that continuously increase without bound as deformation grows. Although these classical models effectively capture nonlinear elasticity and large deformation behaviors, they inherently cannot represent material failure, as real materials eventually rupture when strain energy accumulation approaches a finite limit. Addressing this limitation, Volokh introduced the concept of energy limiters to explicitly model progressive damage and failure within hyperelastic frameworks \cite{Volokh2007,Volokh2010}.

The central idea of the energy limiters approach is to constrain the maximum strain energy density a material element can store, reflecting the physical mechanism of microstructural bond rupture \cite{Volokh2013,Chandrashekar2025}. Specifically, a critical failure energy $\psi_{f}$ (also called saturation energy) is introduced to characterize material toughness, representing the upper bound of strain energy density sustainable by the material before failure occurs. Mathematically, this energy limitation implies that the total strain energy density $\psi$ asymptotically approaches $\psi_{f}$:
\begin{equation}
\lim_{W \to \infty} \psi(W) = \psi_{f},
\end{equation}
where $W$ is the intact (undamaged) hyperelastic strain energy density, a function of applied deformation.

For small-to-moderate deformations, when $W\ll\psi_{f}$, the strain energy density function converges to standard hyperelastic behavior:
\begin{equation}
\psi(W) \approx W, \quad \text{as} \quad W \to 0,
\end{equation}
indicating an initially undamaged response. Conversely, with increasing deformation, the strain energy density approaches the critical failure energy $\psi_{f}$:
\begin{equation}
\psi(W) \to \psi_{f}, \quad \text{as} \quad W \to \infty,
\end{equation}
clearly signifying complete material damage.

\subsection{The Irreducible Tensor Basis Decomposition of Stress in Energy-Limited Hyperelasticity}




The second Piola–Kirchhoff stress tensor $\mathbf{S}$ is derived by differentiating the strain energy density function $\psi$ with respect to the right Cauchy–Green deformation tensor $\mathbf{C}$ \cite{Holzapfel2000}:
\begin{equation}
\mathbf{S} = 2 \frac{\partial \psi(W)}{\partial \mathbf{C}}.
\end{equation}

Using the chain rule, the expression becomes:
\begin{equation}\label{eq:5}
\mathbf{S} = 2 \frac{d\psi}{dW}\frac{\partial W}{\partial \mathbf{C}},
\end{equation}
where the stress-reduction factor $\frac{d\psi}{dW} \in \left[0,1\right]$ reflects progressive material degradation, decreasing monotonically from one (undamaged) to zero (fully damaged) \cite{Chandrashekar2025}.

Assuming isotropic material behavior, the stress tensor can be expressed in terms of scalar invariants of $\mathbf{C}$:
\begin{equation}\label{eq:6}
\mathbf{S} = 2\frac{d\psi}{dW}\frac{\partial W\left( J, I_1, I_2\right)}{\partial \mathbf{C}},
\end{equation}
with the invariants defined as:
\begin{equation}\label{eq:7}
J = \sqrt{\mathrm{det}\mathbf{C}}, \quad I_1= \mathrm{tr}\mathbf{C}, \quad I_2 = \frac{1}{2}\left[ (\mathrm{tr}\mathbf{C})^2 - \mathrm{tr}(\mathbf{C}^2)\right].
\end{equation}

These invariants fully characterize the deformation state for isotropic hyperelastic materials. From a physical perspective, $J$ measures the relative volume change of a material element during deformation (i.e., volumetric dilatation or compression), $I_1$ represents the sum of the squared stretches of an infinitesimal segment averaged over all possible orientations within the material, and $I_2$ is three times the square of the stretch ratio of an infinitesimal area element averaged over all possible orientations, capturing the extent of shear deformation \cite{Kearsley1989,Kilian1985,Currie2004}.

Without loss of generality, $W$ can be decomposed into volumetric and isochoric contributions \cite{Holzapfel2000}:
\begin{equation}\label{eq:8}
W = U(J) + \bar{W}_{iso}\left( \bar{I}_1,\bar{I}_2\right),
\end{equation}
where $U(J)$ describes the volumetric energy density function, and $\bar{W}_{iso}$ characterizes the isochoric strain energy density via invariants of the distortional right Cauchy-Green deformation tensor $\bar{\mathbf{C}} = J^{-2/3}\mathbf{C}$, ensuring volume preservation ($\mathrm{det}\bar{\mathbf{C}}=1$):
\begin{equation}\label{eq:9}
\bar{I}_1= \mathrm{tr}\bar{\mathbf{C}}, \quad \bar{I}_2 = \frac{1}{2}\left[ (\mathrm{tr}\bar{\mathbf{C}})^2 - \mathrm{tr}(\bar{\mathbf{C}}^2)\right].
\end{equation}

Substituting Eq. (\ref{eq:8}) in Eq. (\ref{eq:6}), the total stress can be additively decomposed into volumetric ($\mathbf{S}_{vol}$) and isochoric ($\mathbf{S}_{iso}$) stress components,
\begin{equation}\label{eq:10}
\mathbf{S} = \mathbf{S}_{vol} + \mathbf{S}_{iso} = \frac{d\psi}{dW} \left(2 \frac{\partial U\left(J\right)}{\partial \mathbf{C}} + 2 \frac{\partial \bar{W}_{iso}\left(\bar{I}_1, \bar{I}_2\right)}{\partial \mathbf{C}}\right).
\end{equation}

The partial derivatives in Eq.~(\ref{eq:10}) can be expanded via the chain rule using standard results from isotropic hyperelasticity \cite{Holzapfel2000,Upadhyay2024}, leading to the following Rivlin--Ericksen representation of the total second Piola--Kirchhoff stress:
\begin{equation}\label{eq:11}
\mathbf{S} = \mathbf{S}_{vol} + \mathbf{S}_{iso} = \frac{d\psi}{dW} \left[J \frac{dU}{dJ} \mathbf{C}^{-1} + 2J^{-2/3} \left( \frac{\partial{\bar{W}_{iso}}}{\partial{\bar{I}_1}} + \bar{I}_1\frac{\partial{\bar{W}_{iso}}}{\partial{\bar{I}_2}}\right) \mathrm{Dev}(\mathbf{I}) - 2 J^{-2/3} \frac{\partial{\bar{W}_{iso}}}{\partial{\bar{I}_2}} \mathrm{Dev}\bar{\mathbf{C}}\right],
\end{equation}
%
%
%
%
where $\mathrm{Dev}\left( \cdot\right) = \left(\cdot\right) - \frac{1}{3}\left(\left(\cdot\right):\mathbf{C}\right)\mathbf{C}^{-1}$ is the deviatoric operator in the Lagrangian description, and $\mathbf{I}$ is the identity tensor.

Equation (\ref{eq:11}) is the generalized stress description of the energy-limited hyperelastic constitutive modeling framework. Several phenomenological models, with specific mathematical equations for $\psi(W)$, $U(J)$, and $\bar{W}_{iso}(\bar{I}_1,\bar{I}_2)$ are available in the literature and have been employed to describe the damage responses of various soft materials, including rubber \cite{Xing-gui2022,Mythravaruni2018,Mythravaruni2019,Lev2019}, skin tissue \cite{Pissarenko2020,Li2016}, and artery wall \cite{Volokh2008a,Volokh2011}. Regardless of the choice of these specific constitutive equations, Eq. (\ref{eq:11}) can be rewritten in an alternative form as an additive decomposition of the components of an irreducible tensor bases weighted by certain response functions and modulated by the stress-reduction factor:
\begin{equation}\label{eq:12}
\mathbf{S} = \mathbf{S}_{vol} + \mathbf{S}_{iso} = \raisebox{2pt}{$\chi$}\left( W\right)\left[\zeta(J)\mathbb{G}_1 + J^{-2/3}\Gamma_1(\bar{I}_1, \bar{I}_2)\mathbb{G}_2 + J^{-2/3}\Gamma_2(\bar{I}_1, \bar{I}_2)\mathbb{G}_3\right],
\end{equation}
where the tensor basis components are
\begin{equation}\label{eq:13}
    \mathbb{G}_1 = \mathbf{C}^{-1}, \quad \mathbb{G}_2 = \mathrm{Dev}(\mathbf{I}), \quad \mathbb{G}_3 = \mathrm{Dev}(\bar{\mathbf{C}}),
\end{equation}
the scalar response functions are
\begin{equation} \label{eq:14}
\begin{gathered}
\zeta(J) = J \frac{dU(J)}{dJ}, \\
\Gamma_1(\bar{I}_1, \bar{I}_2) = 2\left( \frac{\partial{\bar{W}_{iso}(\bar{I}_1,\bar{I}_2)}}{\partial{\bar{I}_1}} + \bar{I}_1\frac{\partial{\bar{W}_{iso}(\bar{I}_1,\bar{I}_2)}}{\partial{\bar{I}_2}}\right), \quad
\Gamma_2(\bar{I}_1, \bar{I}_2) = -2 \frac{\partial{\bar{W}_{iso}(\bar{I}_1,\bar{I}_2)}}{\partial{\bar{I}_2}},
\end{gathered}
\end{equation}
and $\raisebox{2pt}{$\chi$} (W) = \frac{d\psi(W)}{dW}$ is the stress-reduction factor.

The generalized stress representation in Eq. (\ref{eq:12}) serves as the foundation for the data-driven constitutive modeling framework proposed in this study. In traditional energy-limited hyperelasticity, response functions (Eq. (\ref{eq:14})) and the stress-reduction factor assume fixed mathematical expressions based on the choices of constitutive equations for $\psi(W)$, $U(J)$, and $W(\bar{I}_1,\bar{I}_2)$ (see Tables 1--3 in \ref{appendix:stress_derivation} for some examples). Unlike these phenomenological models, the data-driven framework proposed in the present study will explicitly learn the $\raisebox{2pt}{$\chi$}$ versus $W$, $\zeta$ versus $J$, and $\{ \Gamma_1, \Gamma_2\}$ versus $\{\bar{I}_1,\bar{I}_2\}$ relationships directly from the experimental stress versus strain data using GPR models, as detailed in the next section.


\section{Proposed Data-Driven Constitutive Modeling Framework}
\label{sec:proposed_modeling_framework}
\subsection{Overview of the Proposed Framework}

\begin{figure}[t]   
    \centering
    \includegraphics[width=5in]{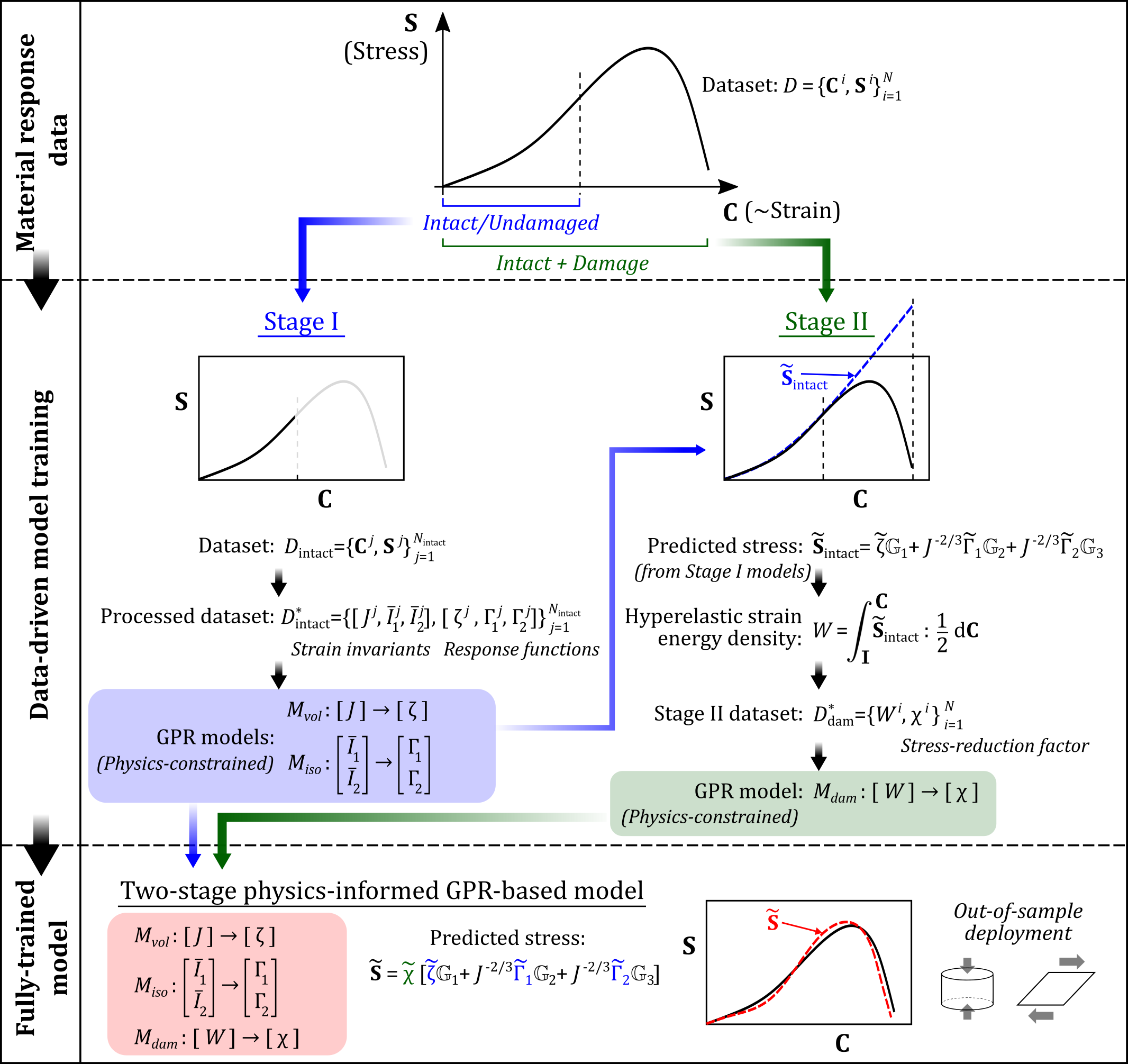}
    \caption{%
    Schematic of the proposed physics-informed data-driven constitutive modeling framework. The complete material response dataset (stress versus strain) comprises both undamaged (intact) and damaged regimes. In Stage I (Section \ref{subsec:stage I model}), the intact portion of the response is processed to compute strain invariants ($J,\bar{I}_1,\bar{I}_2$) and corresponding response functions ($\zeta, \Gamma_1, \Gamma_2$). These quantities serve as the input–output training data for two GPR models that separately characterize the volumetric and isochoric hyperelastic responses, denoted by $\mathcal{M}_{vol}$ and $\mathcal{M}_{iso}$, respectively. In Stage II (Section \ref{subsec:stage II model}), the trained Stage I GPR models are used to predict the intact (i.e., undamaged) stress response and the associated intact hyperelastic strain energy density $W$ over the full loading history. A stress-reduction factor $\raisebox{2pt}{$\chi$}(W)$, representing progressive damage and failure, is then inferred and used to train a third, constrained GPR model $\mathcal{M}_{dam}$. Together, the three GPR models $\mathcal{M}_{vol}$, $\mathcal{M}_{iso}$, and $\mathcal{M}_{dam}$ constitute the proposed \emph{two-stage physics-informed GPR-based model}, enabling physically consistent stress prediction under arbitrary deformation paths.
    }
    \label{fig:modeling_framework}
\end{figure}

The proposed constitutive modeling framework is based on a two-stage GPR approach designed to discover physically interpretable constitutive laws for hyperelastic soft materials undergoing progressive damage. A detailed schematic of this two-stage framework is illustrated in Fig. \ref{fig:modeling_framework}, which outlines the sequence of steps involved in model construction, from data preprocessing to GPR training and final model compilation.

In the first stage, the goal is to characterize the material response in the undamaged regime. This stage operates under the assumption that the mechanical response up to a small-to-intermediate strain level—typically taken to be around 50\% of the maximum strain at complete failure—corresponds to the behavior of the intact material. The exact cutoff is user-defined and can be tuned based on experimental observations. Stress–strain data from this undamaged regime (i.e., when $\raisebox{2pt}{$\chi$}=1$) are used to compute the scalar response functions $\zeta(J)$, $\Gamma_1(\bar{I}_1, \bar{I}_2)$, and $\Gamma_2(\bar{I}_1, \bar{I}_2)$ through inversion of the linearized tensor basis representation of the second Piola–Kirchhoff stress tensor (details in Section \ref{subsec:stage I model}). Two GPR models are then trained: one to learn the mapping between the volumetric strain invariant $J$ and the volumetric response function $\zeta$, and another to learn the mapping between the isochoric invariants $\{\bar{I}_1,\bar{I}_2\}$ and the corresponding isochoric response functions $\{ \Gamma_1, \Gamma_2\}$. Mathematically, these mappings can be expressed as:
\begin{align}
    \begin{split} \label{eq:15}
    \mathcal{M}_{vol}: {} \Bigl[ J \Bigr] &\mapsto \Bigl[ \zeta \Bigr],
    \end{split} \\
    \begin{split} \label{eq:16}
    \mathcal{M}_{iso}: {} \begin{bmatrix}
            \vspace{4 pt}
            \Bar{I}_1 \\
            \Bar{I}_2
         \end{bmatrix} &\mapsto \begin{bmatrix}
            \vspace{4 pt}
            \Gamma_{1} \\
            \Gamma_{2}
         \end{bmatrix}.
    \end{split}
\end{align}

In the second stage, the GPR models obtained from Stage I are used to compute the predicted intact (i.e., undamaged) stress response and the corresponding intact hyperelastic strain energy density $W$ across the full deformation range, including the damaged regime. The stress-reduction factor $\raisebox{2pt}{$\chi$}(W)$ is estimated pointwise from the ratio of the experimental stress to the predicted intact stress using a linear least squares approach (details in Section \ref{subsec:stage II model}). A third GPR model is then trained to learn the mapping from $W$ to $\raisebox{2pt}{$\chi$}$, with physics-based constraints enforcing non-negativity, monotonic decay (i.e., damage can accumulate but not reverse), and an asymptotic behavior (i.e., $\raisebox{2pt}{$\chi$} \rightarrow 0$ as $W \rightarrow \infty$). Mathematically, this mapping is given as:
\begin{equation} \label{eq:17}
    \mathcal{M}_{dam}: {} \Bigl[ W \Bigr] \mapsto \Bigl[ \raisebox{2pt}{$\chi$} \Bigr].
\end{equation}

\emph{Together, the three GPR models, $\mathcal{M}_{vol}$, $\mathcal{M}_{iso}$, and $\mathcal{M}_{dam}$, form a physics-informed data-driven constitutive framework.} Hereafter, we refer to this framework as the \emph{two-stage physics-informed GPR-based model}. Once trained, the complete model can be used to compute the total stress $\mathbf{S}$ for any given input deformation tensor $\mathbf{C}$ as:
\begin{equation} \label{eq:18}
    \widetilde{\mathbf{S}} = \widetilde{\raisebox{2pt}{$\chi$}}\left[\widetilde{\zeta}\mathbb{G}_1 + J^{-2/3}\widetilde{\Gamma}_1\mathbb{G}_2 + J^{-2/3}\widetilde{\Gamma}_2\mathbb{G}_3\right],
\end{equation}
where the accent symbols ( $\widetilde{\cdot}$ ) over the stress, response functions, and stress-reduction factor denote quantities predicted by the constituent GPR models.
%

\subsection{Stage I: Modeling Undamaged Hyperelastic Response}
\label{subsec:stage I model}
\subsubsection{Input--output dataset generation}

In Stage I, an experimental dataset $\mathcal{D}=\{ \mathbf{C}^i, \mathbf{S}^i \}_{i=1}^{N}$, which may include both intact and damaged responses, is first collected. Here, superscript $i$ denotes a particular data point in the dataset, $N$ being the total number of data points. To isolate the intact (undamaged) material response, a user-defined strain (or stretch) cut-off—typically set around 50\% of the maximum strain (or maximum stretch) at failure—is applied. This results in a reduced intact subset $\mathcal{D}_\mathrm{intact}=\{ \mathbf{C}^j, \mathbf{S}^j \}_{j=1}^{N_{\mathrm{intact}}}$, where $N_{\mathrm{intact}}<N$ and $\mathcal{D}_\mathrm{intact} \subset \mathcal{D}$.

From this intact material response subset, an alternative dataset,
\begin{equation} \label{eq:19}
\mathcal{D}_\mathrm{intact}^* = \{[ J^j, \bar{I}_1^j, \bar{I}_2^j ], [\zeta^j, \Gamma_1^j, \Gamma_2^j] \}_{j=1}^{N_{\mathrm{intact}}},
\end{equation}
is generated to facilitate training data-driven mappings. The alternative dataset is created by solving the tensorial linear relationship between the intact stress and tensor basis components (cf. Eq. (\ref{eq:12})),
\begin{equation} \label{eq:20}
    \mathbf{S}_\mathrm{intact} = \zeta\mathbb{G}_1 + J^{-2/3}\Gamma_1\mathbb{G}_2 + J^{-2/3}\Gamma_2\mathbb{G}_3,
\end{equation}
which is expressed in the general linear algebra form ($Ax=b$):
\begin{equation} \label{eq:21}
\underbrace{
\begin{bmatrix}
\mathrm{vec}(\mathbb{G}_1) & J^{-2/3}\mathrm{vec}(\mathbb{G}_2) & J^{-2/3}\mathrm{vec}(\mathbb{G}_3)
\end{bmatrix}}_{A}
\underbrace{
\begin{bmatrix}
\zeta \\
\Gamma_1 \\
\Gamma_2
\end{bmatrix}}_{x}
=
\underbrace{\text{vec}(\mathbf{S}_{\mathrm{intact}})}_{b},
\end{equation}
where $\mathrm{vec}(\cdot) \in \mathbb{R}^{9\times1}$ denotes the vectorization of tensors. This equation is solved at each experimental data point in $\mathcal{D}_\mathrm{intact}$---with known values of $\mathbf{S}_\mathrm{intact}$, $J$, $\mathbb{G}_1$, $\mathbb{G}_2$, and $\mathbb{G}_3$ (latter four computed from $\mathbf{C}$ using Eqs. (\ref{eq:7}) and (\ref{eq:13}))---to calculate the three response functions by minimizing the residual: $\min_{\zeta,\Gamma_1,\Gamma_2} \left\| A x - b \right\|^2$. Note that the known values of $\mathbf{S}_\mathrm{intact}$ correspond to the stress measurements contained in the reduced intact dataset $\mathcal{D}_\mathrm{intact}$.

Alternatively, if elastic volumetric and isochoric responses are available separately (e.g., from independent volumetric compression and isochoric deformation tests), the response functions may be obtained by solving two decoupled linear systems of the form $Ax=b$. In this case, the volumetric response function $\zeta$ can be extracted by solving an equation involving $\mathrm{vec}(\mathbf{S}_{\mathrm{intact},vol})$ and the vectorized tensor basis $\mathrm{vec}(\mathbb{G}_1)$, while the isochoric response functions $\Gamma_1$ and $\Gamma_2$ can be obtained from $\mathrm{vec}(\mathbf{S}_{\mathrm{intact},iso})$ and $\left(J^{-2/3}\mathrm{vec}(\mathbb{G}_2), J^{-2/3}\mathrm{vec}(\mathbb{G}_3)\right)$. This alternative procedure yields the same sets of data points $\{\zeta^j, \Gamma_1^j, \Gamma_2^j\}$ used to construct the dataset $\mathcal{D}_\mathrm{intact}^*$.

Once the dataset $\mathcal{D}^*_\mathrm{intact}$ is created, the two GPR models $\mathcal{M}_{vol}:J\mapsto\zeta$ (Eq. (\ref{eq:15})) and $\mathcal{M}_{iso}:(\bar{I}_1,\bar{I}_2)\mapsto(\Gamma_1,\Gamma_2)$ (Eq. (\ref{eq:16})) are trained separately. This completes Stage I, establishing a robust, physically consistent baseline model for undamaged hyperelastic response predictions.

\subsubsection{Gaussian process regression implementation}
In this work, GPR is employed as a nonparametric Bayesian regression framework to learn functional relationships directly from data \cite{GPR1995}. Given input--output pairs $\{\mathbf{z}^j,\mathbf{y}^j\}_{j=1}^{n}$, the outputs are modeled as realizations of a zero-mean Gaussian process:
\begin{equation} \label{eq:22}
\mathbf{y}(\mathbf{z}) \sim \mathcal{GP}(0, k(\mathbf{z},\mathbf{z}')),
\end{equation}
where $\mathbf{z} \in \mathbb{R}^{d_{\mathrm{in}}}$ denotes the input vector (e.g., $J$ and $[\bar{I}_1, \bar{I}_2]^\mathrm{T}$), and $\mathbf{y}(\mathbf{z}) \in \mathbb{R}^{d_{\mathrm{out}}}$ is the corresponding vector-valued output (e.g., $\zeta$ or $[\Gamma_1,\Gamma_2]^\mathrm{T}$). For the volumetric GPR model $\mathcal{M}_{vol}$, $d_{\mathrm{in}} = d_{\mathrm{out}} = 1$, whereas for the isochoric GPR model $\mathcal{M}_{iso}$, $d_{\mathrm{in}} = d_{\mathrm{out}} = 2$.

The covariance kernel $k(\mathbf{z},\mathbf{z}')$ encodes prior assumptions regarding the smoothness, correlation length, and regularity of the underlying function being learned. In essence, the kernel determines how outputs at two different input locations are statistically correlated, thereby governing the functional class that the GPR model can represent. We employ the Matérn 5/2 covariance kernel \cite{Stein1999,Matern1986}, augmented with a nugget term for numerical stability:
\begin{equation} \label{eq:23}
    k(\mathbf{z},\mathbf{z}') = \sigma_f^2\left(1+\frac{\sqrt{5}r}{\ell}+\frac{5r^2}{3\ell^2}\right)\exp\left(-\frac{\sqrt{5}r}{\ell}\right) + \alpha\delta_{\mathbf{z}\mathbf{z}'},
\end{equation}
where $r=\left\Vert\mathbf{z}-\mathbf{z}'\right\Vert$ is the Euclidean distance, $\sigma_f^2$ is the signal variance, $l$ is the characteristic length scale, $\alpha$ is the nugget parameter, and $\delta_{\mathbf{z}\mathbf{z}}'$ is the Kronecker delta.

The Matérn 5/2 kernel is chosen based on the comparative study by Laurent et al. \cite{Laurent2019}, which demonstrated its robustness for computer experiments when little prior information is available about the underlying functional form. Importantly, this kernel offers a controlled level of smoothness—corresponding to twice mean-square differentiability—making it sufficiently flexible for constitutive modeling while avoiding the overly smooth behavior associated with squared-exponential kernels \cite{Thant2025,Palar2023}. Note, kernel differentiability will be particularly critical for the Stage II damage model (Section \ref{subsec:stage II model}), where constraints are imposed directly on the derivative of the learned function with respect to the strain energy density.

The optimal kernel hyperparameters $\boldsymbol{\theta}=[\sigma_f,l]$ that best describe the training dataset are obtained by minimizing the negative marginal log-likelihood \cite{GPR2003}:
\begin{equation} \label{eq:24}
\widetilde{{\boldsymbol{\theta}}} = \argmin_{ {\boldsymbol{\theta}}^{\star}} \left[-\log \, p(\mathbf{Y} | \mathbf{Z}, \boldsymbol{\theta})\right] =  \argmin_{ {\boldsymbol{\theta}}^{\star}} \left[\frac{1}{2} \mathbf{Y}^{\mathrm{T}} \mathbf{K}(\mathbf{Z},\mathbf{Z}) ^{-1} \mathbf{Y} + \frac{1}{2} \text{log}( \det ( \mathbf{K}(\mathbf{Z},\mathbf{Z}) )) + \frac{n}{2} \log (2 \pi) \right],
\end{equation}
where the matrix $\mathbf{Z} \in \mathbb{R}^{n\times d_{\mathrm{in}}}$ denotes the full set of input training data, with each row representing a sample data point; similarly, $\mathbf{Y} \in \mathbb{R}^{n\times d_{\mathrm{out}}}$ contains all vector-valued outputs in the training dataset. $\mathbf{K}(\mathbf{Z},\mathbf{Z})$ is the Gram matrix assembled using the kernel $k(\cdot,\cdot)$.

After finding the best parameters, a GPR regression model is fully defined. Given a new input data point $\mathbf{z}_{\star}$, the predicted output $\widetilde{\mathbf{y}}$ of the Gaussian process regressor reads
\begin{equation}\label{eq:25}
\begin{aligned}
\widetilde{\mathbf{y}}({\mathbf{z}}_{\star}) &=   \mathbf{K}(\mathbf{Z},\mathbf{z}_{\star})^{\mathrm{T}} \, \mathbf{K}(\mathbf{Z},\mathbf{Z}) ^{-1} \, \mathbf{Y},
\end{aligned}
\end{equation}
with predictive variance
\begin{equation}\label{eq:26}
\begin{aligned}
\widetilde{\sigma}^{2}({\mathbf{z}}_{\star}) &= k(\mathbf{z}_{\star},\mathbf{z}_{\star}) -  \mathbf{K}(\mathbf{Z},\mathbf{z}_{\star})^\mathrm{T} \, \mathbf{K}(\mathbf{Z},\mathbf{Z}) ^{-1}\,  \mathbf{K}(\mathbf{Z},\mathbf{z}_{\star}).
\end{aligned}
\end{equation}

As $\alpha \rightarrow 0$ in Eq. (\ref{eq:23}), the GPR predictor exhibits the exact inference property, i.e., $\widetilde{\mathbf{y}}(\mathbf{z}^{j})= \mathbf{ y}^{j}$ at all training points $j=1, \ldots, n$. This property is exploited to enforce the stress-free reference state by explicitly including the data point $\{\mathbf{C}=\mathbf{I}, \mathbf{S}=\mathbf{0}\}$ in the training set. In terms of invariants, $\mathbf{C}=\mathbf{I}$ corresponds to $J=1$ and $\Bar{I}_1 = \Bar{I}_2 = 3$, while $\mathbf{S}=\mathbf{0}$ translates to $\zeta=\Gamma_1=\Gamma_2=0$.

Hyperparameter optimization is performed using the Limited-memory Broyden–Fletcher–Goldfarb– Shanno (L-BFGS-B) algorithm in the SciPy library for Python \cite{2020SciPy-NMeth}.

\subsection{Stage II: Incorporating Progressive Damage and Failure with Constrained GPR}
\label{subsec:stage II model}

\subsubsection{Input--output dataset generation}

In Stage II, the stress-reduction factor $\raisebox{2pt}{$\chi$} $, which characterizes material degradation, is modeled explicitly. Using the GPR models $\mathcal{M}_{vol}$ and $\mathcal{M}_{iso}$ trained in Stage I, the intact stress $\widetilde{\mathbf{S}}_\mathrm{intact}$ is predicted across the entire deformation range in the training dataset $\mathcal{D}$ (including the damaged region) as
\begin{equation} \label{eq:26.5}
    \widetilde{\mathbf{S}}_\mathrm{intact} = \widetilde{\zeta}\mathbb{G}_1 + J^{-2/3}\widetilde{\Gamma}_1\mathbb{G}_2 + J^{-2/3}\widetilde{\Gamma}_2\mathbb{G}_3,
\end{equation}
where the accent symbols ( $\widetilde{\cdot}$ ) denote predicted values from Stage I GPR models.

The corresponding intact hyperelastic strain energy density $W$ at each data point is calculated using
\begin{equation} \label{eq:27}
W = \int_{\mathbf{I}}^{\mathbf{C}} \widetilde{\mathbf{S}}_{\mathrm{intact}} : \frac{1}{2}\,\mathrm{d}\mathbf{C}.
\end{equation}

Numerically, this integration is performed using the trapezoidal rule, which balances simplicity and accuracy and is particularly suitable when the stress and deformation tensors are available at finely sampled intervals. Using Eq. (\ref{eq:27}), a $W$ (i.e., $W^i$ for $i \in [1,N]$) is computed for each $\mathbf{C}^i$ in $\mathcal{D}$, creating the input dataset for the Stage II model. Each computation of integral utilizes the full discrete sequence of deformation states from $\mathbf{C}^1=\mathbf{I}$ to $\mathbf{C}^i$.

The stress-reduction factor at each data point is then determined by solving the linear relationship (see Eq. (\ref{eq:12})):
\begin{equation} \label{eq:28}
\mathbf{S} = \raisebox{2pt}{$\chi$}\widetilde{\mathbf{S}}_{\mathrm{intact}},
\end{equation}
which, when expressed in vectorized form, leads to the following linear system:
\begin{equation} \label{eq:29}
\underbrace{\text{vec}(\widetilde{\mathbf{S}}_{\mathrm{intact}})}_{A}\,\underbrace{\chi}_{x} = \underbrace{\text{vec}(\mathbf{S})}_{b}.
\end{equation}

This equation is solved via least squares minimization: $\min_{\chi}\|A x - b\|^2$.

Using $W$ and $\raisebox{2pt}{$\chi$} $ calculated at each training data point, an alternative training dataset,
\begin{equation} \label{eq:29.5a}
\mathcal{D}^*_\mathrm{dam}=\{ W^i, \raisebox{2pt}{$\chi$}^i \}_{i=1}^{N},
\end{equation}
is thereby constructed to train a third GPR model $\mathcal{M}_{dam}:W \mapsto \raisebox{2pt}{$\chi$} $ (Eq. (\ref{eq:17})). Unlike the Stage I GPR models, the Stage II model is a constrained GPR regressor that explicitly incorporates physics-based constraints to ensure realistic damage evolution behavior.

\subsubsection{Constrained Gaussian process regression for damage modeling}
\label{subsubsec:3.3.2}

To ensure physically admissible damage evolution, the GPR model for the stress-reduction factor, $\mathcal{M}_{dam}$, is trained using a constrained hyperparameter optimization strategy that enforces three essential physical requirements: (i) non-negativity of the stress-reduction factor, (ii) monotonic damage accumulation, and (iii) complete material failure at large strains (i.e, at large $W$). Unlike the Stage I models, which are trained using standard minimum negative marginal log-likelihood estimation, the Stage II model incorporates these constraints through penalty-based regularization of the negative marginal log-likelihood, evaluated at selected constraint points.
\begin{enumerate}
  \item \emph{Non-negativity of the stress-reduction factor:} Physical admissibility requires the stress-reduction factor to satisfy
  \begin{equation} \label{eq:29.5b}
    \raisebox{2pt}{$\chi$}(W) \geq 0 \quad \forall W.
  \end{equation}

  To enforce this condition, a non-negativity penalty is introduced at a set of constraint inputs, i.e., selected values of the intact strain energy density used exclusively for constraint enforcement:
  \begin{equation} \label{eq:29.5c}
    \mathcal{D}_{\mathrm{cons}}=\{W_c^p\}_{p=1}^{N_c}.
  \end{equation}
  Violations of non-negativity are penalized through the following quadratic penalty term:
  \begin{equation} \label{eq:29.5d}
    \mathcal{P}_\mathrm{nn}(\boldsymbol{\theta})= \Lambda_\mathrm{nn}\sum_{p=1}^{N_c} {\left[\mathrm{min}\left( 0, \mathbb{E} \left[\raisebox{2pt}{$\chi$}(W^p_c)\right]\right)\right]^2},
  \end{equation}
  where $\mathbb{E} \left[\raisebox{2pt}{$\chi$}(W^p_c)\right]$ denotes the posterior mean of the GPR prediction evaluated at constraint points, and $\Lambda_\mathrm{nn}$ is a user-defined penalty weight.
  
  \item \emph{Monotonicity of the stress-reduction factor:} To enforce irreversible damage accumulation, the stress-reduction factor must be a monotonically decreasing (i.e., non-increasing) function of the intact hyperelastic strain energy density:
  \begin{equation} \label{eq:30}
    \frac{d\raisebox{2pt}{$\chi$}}{dW} \leq 0.
  \end{equation}

  Since derivatives of a Gaussian process are themselves Gaussian processes, the derivative $\frac{d\raisebox{2pt}{$\chi$}}{dW}$ can be evaluated analytically through derivatives of the covariance kernel. Specifically, for the zero-mean GPR model $\mathcal{M}_{dam}$, we have \cite{Riihimaki2010}
  \begin{equation} \label{eq:30.5A}
    \raisebox{2pt}{$\chi$}(W) \sim \mathcal{GP}\left(0, k(W,W')\right) \implies \frac{d\raisebox{2pt}{$\chi$}}{dW} \sim \mathcal{GP}\left(0, \frac{\partial^2 k(W,W')}{\partial W \partial W'}\right).
  \end{equation}

  The posterior mean of the derivative at a constraint point $W^p_c$ is given by \cite{Riihimaki2010}:
  \begin{equation} \label{eq:31}
      \mathbb{E}\left[\at{\frac{\partial \raisebox{2pt}{$\chi$}}{\partial W}}{W=W_c^p}\right] = \frac{\partial \mathbf{K}(\boldsymbol{W},W_c^p)}{\partial W'}^\mathrm{T}\mathbf{K}(\boldsymbol{W},\boldsymbol{W})^{-1}\boldsymbol{\raisebox{2pt}{$\chi$}},
  \end{equation}
  where $\boldsymbol{W} = [W^{1}, \ldots, W^{N}]^\mathrm{T}$ denotes the vector of all Stage II training inputs from the dataset $\mathcal{D}^*_\mathrm{dam}$, and $\boldsymbol{\raisebox{2pt}{$\chi$}} = [\raisebox{2pt}{$\chi$}(W^{1}), \ldots, \raisebox{2pt}{$\chi$}(W^{N})]^\mathrm{T} = [\raisebox{2pt}{$\chi$}^{1}, \ldots, \raisebox{2pt}{$\chi$}^{N}]^\mathrm{T}$ are the corresponding training outputs. Here, $k(\cdot,\cdot)$ denotes the scalar covariance kernel and $\mathbf{K}(\cdot,\cdot)$ the associated Gram matrix.

  For the Matérn 5/2 kernel $k(W,W')$ used in this work (Eq. (\ref{eq:23})), the kernel derivative is:
  \begin{equation} \label{eq:33}
    \frac{\partial k(W,W')}{\partial W'} = \sigma_f^2 \left( \frac{5(W-W')}{3\ell^2} \right) \left(1+\frac{\sqrt{5}r}{\ell}\right) \exp\left(-\frac{\sqrt{5}r}{\ell}\right),
  \end{equation}
  with $r=|W-W'|$.

  To discourage violations of monotonicity, a quadratic penalty is imposed on positive derivative values (evaluated at data points in $\mathcal{D}_\mathrm{cons}$):
  \begin{equation} \label{eq:34}
    \mathcal{P}_\mathrm{mono}(\boldsymbol{\theta})= \Lambda_\mathrm{mono}\sum_{p=1}^{N_c} {\left[\mathrm{min}\left( 0, -\mathbb{E}\left[\at{\frac{\partial \raisebox{2pt}{$\chi$}}{\partial W}}{W=W_c^p}\right]\right)\right]^2},
  \end{equation}
  where $\Lambda_\mathrm{mono}$ is a user-defined penalty weight. This formulation enforces monotonic decay of the expected damage evolution, while retaining numerical robustness and compatibility with standard optimization algorithms.

  \item \emph{Complete failure at very large strains:} To represent material failure, the stress-reduction factor must asymptotically approach zero as the intact hyperelastic strain energy density becomes large:
  \begin{equation} \label{eq:35}
    \lim_{W \to \infty}{\raisebox{2pt}{$\chi$} (W) = 0}.
  \end{equation}

  Rather than enforcing this behavior analytically, it is incorporated through data augmentation. Specifically, additional artificial points are introduced in the training dataset $\mathcal{D}^*_\mathrm{dam}$ in the range $W \in [aW_\mathrm{peak},bW_\mathrm{peak}]$, where $a$ and $b$ ($b>a>1$) are adjustable constants and $W_\mathrm{peak}$ is the maximum intact hyperelastic strain energy density value in the original dataset $\mathcal{D}^*_\mathrm{dam}$ prior to data augmentation. At these artificial points, $\raisebox{2pt}{$\chi$}(W)=0$ is prescribed. 
\end{enumerate}

Combining the likelihood term with the penalty-based constraints, the final optimization problem for the Stage II GPR hyperparameters $\boldsymbol{\theta}$ is formulated as:
\begin{equation} \label{eq:35.5}
\widetilde{{\boldsymbol{\theta}}} = \argmin_{ {\boldsymbol{\theta}}^{\star}} \left[-\log \, p(\boldsymbol{\raisebox{2pt}{$\chi$}} | \boldsymbol{W}, \boldsymbol{\theta}) + \mathcal{P}_\mathrm{nn}(\boldsymbol{\theta}) + \mathcal{P}_\mathrm{mono}(\boldsymbol{\theta})\right],
\end{equation}
where the augmented dataset implicitly enforces the complete failure constraint. The constrained optimization problem is solved using the L-BFGS-B algorithm in the SciPy library \cite{2020SciPy-NMeth}.

Once trained, the three GPR models—$\mathcal{M}_{vol}$, $\mathcal{M}_{iso}$, and $\mathcal{M}_{dam}$—constitute the complete two-stage physics-informed GPR-based model. For an arbitrary deformation tensor $\mathbf{C}$, the Stage I models $\mathcal{M}_{vol}$ and $\mathcal{M}_{iso}$ predict the volumetric and isochoric hyperelastic response functions, $\widetilde{\zeta}$, $\widetilde{\Gamma}_1$ and $\widetilde{\Gamma}_2$, which are used to reconstruct the intact stress response and the corresponding intact hyperelastic strain energy density. The Stage II model $\mathcal{M}_{dam}$ then predicts the evolution of the stress-reduction factor $\widetilde{\raisebox{2pt}{$\chi$}}$, capturing progressive material degradation and failure. The combination of predicted response functions and stress-reduction factor evolution enables prediction of the complete stress–strain response using Eq. (\ref{eq:18}).
%

\subsection{Physics-Based Constraints Satisfied by the Proposed Framework}

The proposed two-stage physics-informed GPR-based model inherently satisfies several fundamental physical principles required of continuum constitutive models, as discussed comprehensively in Tadmor et al.~\cite{Tadmor2011}. These constraints are summarized below:

\begin{itemize}
    \item \emph{Principle of Determinism:} The model establishes a deterministic mapping between the input deformation tensor $\mathbf{C}$ and the output stress tensor $\mathbf{S}$. For any given $\mathbf{C}$, the model predicts a unique, reproducible stress response through Eq.~(\ref{eq:18}).

    \item \emph{Principle of Local Action:} The model uses strain invariants $J$, $\bar{I}_1$, and $\bar{I}_2$ derived from the local deformation tensor $\mathbf{C}$. Consequently, the stress at a material point (Eq.~(\ref{eq:18})) depends only on the local deformation, not on neighboring points.

    \item \emph{Balance of Angular Momentum:} Since the tensor basis components (Eq.~(\ref{eq:13})) are symmetric and the stress tensor is expressed as their linear combination (Eq.~(\ref{eq:18})), the symmetry of the second Piola--Kirchhoff stress tensor $\mathbf{S}$ is automatically satisfied, i.e., $\mathbf{S} = \mathbf{S}^\mathrm{T}$, thereby fulfilling the local balance of angular momentum.

    \item \emph{Material Frame-Indifference (Objectivity):} Material frame-indifference is ensured because $\mathbf{S}$ and $\mathbf{C}$ are both objective tensors associated with the reference configuration and thus remain unaffected under superimposed rigid body motions. In other words, $\mathbf{S}(\mathbf{C}) = \mathbf{R}^T \mathbf{S}(\mathbf{R} \mathbf{C} \mathbf{R}^T) \mathbf{R} \quad \forall \mathbf{R} \in SO(3)$, where $SO(3)$ denotes the special orthogonal group of proper rotations in three dimensions..

    \item \emph{Isotropic Material Symmetry:} Isotropy follows directly from the generalized constitutive formulation in Eq.~(\ref{eq:6}), which depends solely on the isotropic invariants of $\mathbf{C}$: $J$, $\bar{I}_1$, and $\bar{I}_2$.

    \item \emph{Stress-Free Reference State (Normalization Condition):} As described in Stage I (Section~\ref{subsec:stage I model}), the stress-free undeformed configuration is enforced by explicitly including the data point corresponding to zero strain and zero stress ($\mathbf{C} = \mathbf{I}, \mathbf{S} = \mathbf{0}$) in the training dataset. Owing to the exact inference property of GPR with a small nugget value, the trained model recovers zero stress at the reference state exactly.

    \item \emph{Thermodynamic Consistency:} The intact (undamaged) stress response learned in Stage I is inherently thermodynamically consistent, as it is derived from the intact hyperelastic strain energy density $W$ via differentiation with respect to $\mathbf{C}$ (Eq. (\ref{eq:5})), consistent with the Coleman–Noll procedure \cite{Hutter2018}. As enforced in Stage II (Section~\ref{subsec:stage II model}), the stress-reduction factor $\raisebox{2pt}{$\chi$}(W)$ is further constrained to be a non-negative, monotonically decreasing function of $W$, i.e., $\raisebox{2pt}{$\chi$} \geq 0$ and $\frac{d\raisebox{2pt}{$\chi$}}{dW} \leq 0$. This ensures that the mechanical work supplied to the system is either stored elastically (up to a saturation limit) or dissipated through internal damage, and never leads to spontaneous energy creation. Additional synthetic data points beyond the maximum observed $W$ further enforce the asymptotic condition $\lim_{W \to \infty}{\raisebox{2pt}{$\chi$} (W) = 0}$, representing complete material failure and maximal internal degradation.
\end{itemize}

The monotonic decay of $\raisebox{2pt}{$\chi$}(W)$ induces progressive material softening with increasing deformation, manifested as a reduction in the material tangent stiffness. Beyond peak load, the stress--strain response may exhibit a negative slope, reflecting material instability and the onset of failure. In this regime, localized stiffness degradation can result in loss of ellipticity, characterized by negative eigenvalues of the \emph{acoustic tensor}, which is a second-order tensor derived from the fourth-order tangent stiffness tensor. Such behavior is physically consistent with the breakdown of material integrity in energy-limited hyperelastic models (see Volokh~\cite{Volokh2017}). A complete derivation of the tangent stiffness tensor, which is essential for numerical implementation (e.g., finite element formulations), is provided in ~\ref{appendix:stiffness}.
%

\section{Model Evaluation Methodology}
\label{sec:Model_eval_method}

This section outlines the datasets and evaluation procedures used to assess the predictive performance of the proposed two-stage physics-informed GPR-based model. Two studies are performed: (i) validation against numerically generated synthetic datasets with known ground-truth constitutive behavior, and (ii) application to modeling experimental uniaxial tension data of a real soft biological tissue. Unless stated otherwise, model comparisons are carried out using the stress-based relative error metric defined in Section~\ref{sec:4.3}. 

To maintain focus on dataset design and evaluation protocol, implementation-specific choices used in GPR training (e.g., nugget values, stretch cutoffs for isolating intact response, constraint point selection, penalty weights, and the range/number of artificially augmented failure points) are not enumerated here; these settings are instead reported alongside the corresponding fits and predictions in the Results and Discussion section (Section \ref{sec:results_and_discussion}).

\subsection{Validation Using Synthetic Data}
\label{subsec:4.1}

To systematically assess accuracy and out-of-sample generalizability, synthetic datasets are generated by combining the Simo--Miehe volumetric energy density function \cite{Simo1992} and the Mooney--Rivlin isochoric strain energy density function \cite{Mooney1940,Rivlin1948b} with Volokh's universal energy-limiter damage formulation \cite{Volokh2010}. The total strain energy density is given by
\begin{equation} \label{eq:36}
    \psi = \frac{\Phi}{m}\left\{ \mathit{\Gamma^*}\left( \frac{1}{m},0 \right) - \mathit{\Gamma^*}\left( \frac{1}{m}, \left(\frac{W}{\Phi}\right)^m \right)\right\},
\end{equation}
where
\begin{equation} \label{eq:37}
    W = U(J) + \bar{W}_{iso}(\bar{I}_1,\bar{I}_2)
    = \frac{\kappa}{2}\left(\frac{J^2-1}{2} - \ln J\right)
    + A_{10}(\bar{I}_1-3) + A_{01}(\bar{I}_2-3),
\end{equation}
and the upper incomplete gamma function is
\begin{equation} \label{eq:38}
    \mathit{\Gamma^*}(s,x)=\int_x^\infty{t^{s-1}\exp(-t)}dt.
\end{equation}
Here, $\Phi$ is the pseudo-failure energy, $m$ is the softening sharpness parameter, $\kappa$ is the bulk modulus, and $A_{10}$, $A_{01}$ are Mooney--Rivlin material parameters. The corresponding critical failure energy is $\psi_{f}=\frac{\Phi}{m}\mathit{\Gamma^*}\left(\frac{1}{m}, 0\right)$ \cite{Volokh2010}.

\begin{figure}[t]
    \centering
    \includegraphics[width=5in]{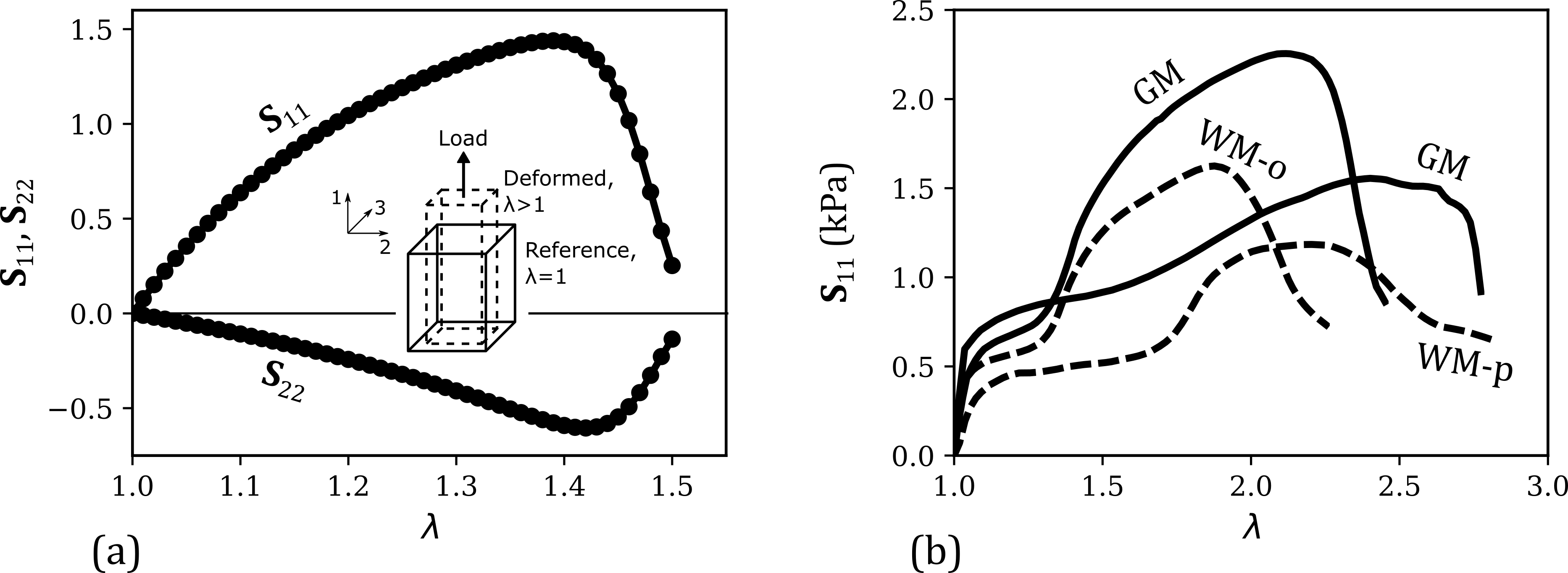}
    \caption{%
    (a) Second Piola–Kirchhoff stress components $\mathbf{S}_{11}$ and $\mathbf{S}_{22}$ versus stretch $\lambda$ for the synthetic uniaxial tension dataset used in numerical validation. Inset shows a schematic illustration of the uniaxial tension deformation, highlighting the reference and deformed states and the 11-loading direction. (b) Axial second Piola–Kirchhoff stress $\mathbf{S}_{11}$ versus $\lambda$ for brain tissue specimens reported by Franceschini et al. \cite{Franceschini2006} and used to demonstrate model application to real experimental data. Solid curves denote gray matter (GM) of thalamus; dashed curves correspond to white matter from the occipital (WM-o) and parietal (WM-p) lobes.
    }
    \label{fig:training_framework}
\end{figure}

The training dataset $\mathcal{D}=\{ \mathbf{C}^i, \mathbf{S}^i \}_{i=1}^{N}$ (with $N=51$) is generated by simulating uniaxial tensile deformations using the following deformation gradient tensor:
\begin{equation} \label{eq:39}
    \mathbf{F} = \mathrm{diag}\left( \lambda, \lambda^{-\nu}, \lambda^{-\nu} \right), 
    \quad \lambda \in \left[1,1.5\right],
\end{equation}
where \( \lambda \) is the axial stretch and \( \nu \) is the Poisson’s ratio. The corresponding right Cauchy--Green tensor is
\begin{equation} \label{eq:40}
\mathbf{C} = \mathbf{F}^\mathrm{T} \mathbf{F} 
= \mathrm{diag}\left( \lambda^2, \lambda^{-2\nu}, \lambda^{-2\nu} \right).
\end{equation}
In this study, a nearly incompressible response is approximated using \( \nu = 0.49 \). The second Piola--Kirchhoff stress tensor \(\mathbf{S}\) at each deformation state is computed using Eq.~(\ref{eq:11}) as
\begin{equation} \label{eq:41}
    \mathbf{S} = \exp\left(-\left(\frac{W}{\Phi}\right)^m\right)
    \left[\frac{\kappa}{2} \left( J^2 - 1\right) \mathbf{C}^{-1}
    + J^{-2/3}\left[2(A_{10} + \bar{I}_1A_{01})\mathrm{Dev}\mathbf{I}
    - 2A_{01}\mathrm{Dev}\bar{\mathbf{C}}\right]\right],
\end{equation}
using parameter values $\Phi=0.75$, $m=10$, $\kappa=100$, $A_{10}=1$, and $A_{01}=0.5$. Figure \ref{fig:training_framework}(a) shows the resulting uniaxial tension responses in terms of the nonzero stress components $\mathbf{S}_{11}$ and $\mathbf{S}_{22}$ (with $\mathbf{S}_{22}=\mathbf{S}_{33}$). 

After training the proposed data-driven model on uniaxial tension data, out-of-sample generalization is evaluated under deformation modes not included in the training set: (i) uniaxial compression with \(\lambda \in [0.5,1]\), and (ii) simple shear, characterized by
\begin{equation} \label{eq:42}
\mathbf{F} = \begin{bmatrix}
1 & \gamma & 0 \\
0 & 1 & 0 \\
0 & 0 & 1
\end{bmatrix},
\end{equation}
where \(\gamma \in [0,0.8]\) is the shear strain magnitude.

The predictive performance of the proposed model is benchmarked against:
\begin{itemize}
    \item A theory-based analytical constitutive model fit to the training data, combining a neo-Hookean volumetric energy density \cite{DeRooij2016} and a two-parameter Yeoh isochoric strain energy density \cite{Yeoh1990,Yeoh1993} with Volokh’s one-parameter reduced damage formulation \cite{Volokh2007}:
    \begin{equation} \label{eq:43}
    \psi = \Phi - \Phi\exp\left(-\frac{W}{\Phi}\right), 
    \quad \text{with} \quad
    W = \frac{\kappa}{2}\left(J-1\right)^2 + C_1(\bar{I}_1-3) + C_2(\bar{I}_1-3)^2.
    \end{equation}
    The calibrated model parameters are $\kappa = 99.28$, $C_1 = 1.73$, $C_2 = -0.55$, and $\Phi = 3.05$.
    \item A black-box GPR model directly mapping components of \(\mathbf{C}\) to components of \(\mathbf{S}\):
    \begin{equation} \label{eq:44}
    \mathcal{M}_{black-box}: \ \Bigl[ \mathrm{vec}(\mathbf{C}) \Bigr] \mapsto \Bigl[ \mathrm{vec}(\mathbf{S}) \Bigr].
    \end{equation}
    \item A single-stage (direct) GPR-based model mapping invariants \((J, \bar{I}_1, \bar{I}_2)\) to \emph{scaled} response functions that appear in the stress decomposition (Eq. (\ref{eq:12})):
    \begin{align}
    \begin{split} \label{eq:45}
    \mathcal{M}_{direct,vol}: {} \Bigl[ J \Bigr] &\mapsto \Bigl[ \raisebox{2pt}{$\chi$}\zeta \Bigr],
    \end{split} \\
    \begin{split} \label{eq:46}
    \mathcal{M}_{direct,iso}: {} \begin{bmatrix}
            \vspace{4 pt}
            \Bar{I}_1 \\
            \Bar{I}_2
         \end{bmatrix} &\mapsto \begin{bmatrix}
            \vspace{4 pt}
            \raisebox{2pt}{$\chi$}\Gamma_{1} \\
            \raisebox{2pt}{$\chi$}\Gamma_{2}
         \end{bmatrix}.
    \end{split}
    \end{align}
    Scaled response functions are obtained by applying the same invariant computation and tensor basis stress decomposition used in the proposed framework (Eq.~(\ref{eq:12})), but fitting GPR mappings directly to the combined (intact + damaged) data in a single step so that the learned targets are the multiplicative coefficients appearing in the stress representation, i.e., $\raisebox{2pt}{$\chi$} \zeta(J)$ and $\raisebox{2pt}{$\chi$} \Gamma_{1,2}(\bar{I}_1,\bar{I}_2)$.
\end{itemize}
\noindent Quantitative comparisons are performed using the relative error metric detailed in Section~\ref{sec:4.3}.
%

\subsection{Application to Soft Tissue Mechanical Response}
\label{sec:4.2}

To assess applicability to experimental soft-tissue behavior, the proposed model is applied to capturing uniaxial tension data of brain tissue gray and white matter reported by Franceschini et al.~\cite{Franceschini2006}. The original dataset comprises nominal (first Piola--Kirchhoff) stress $\mathbf{P}_{11}$ versus stretch $\lambda$ measurements under monotonic tension to very large stretches (exceeding 2), exhibiting progressive softening and eventual failure. Figure \ref{fig:training_framework}(b) shows the corresponding axial second Piola--Kirchhoff stress component versus stretch responses, computed using $\mathbf{S}_{11}=\mathbf{P}_{11}/(1+\lambda)$.

\paragraph{\textbf{Training with uniaxial-only data under incompressibility constraint}}
Unlike the synthetic validation, the experimental dataset does not include volumetric or hydrostatic loading information. Therefore, consistent with common practice for many soft materials, the material response is assumed incompressible ($J=1$) \cite{Budday2020,Liu2025b,Chockalingam2020,Upadhyay2021,Upadhyay2019b,Upadhyay2020b}. Under incompressibility, the volumetric contribution is replaced by a Lagrange multiplier $p$ enforcing the constraint, and the stress decomposition no longer involves an independently modeled volumetric response function $\zeta(J)$. Consequently, the Stage I training procedure is modified: only the isochoric response functions are extracted and learned (i.e., only $\mathcal{M}_{iso}$ is trained in Stage I), followed by Stage II damage learning as in the general framework.

Under incompressibility, the second Piola--Kirchhoff stress tensor takes the form (cf. Eq. (\ref{eq:12}))
\begin{equation} \label{eq:47}
    \mathbf{S} =  -p\,\mathbb{G}_1 
    + \raisebox{2pt}{$\chi$}\left( W\right)\left[\Gamma_1(\bar{I}_1, \bar{I}_2)\mathbb{G}_2 
    + \Gamma_2(\bar{I}_1, \bar{I}_2)\mathbb{G}_3\right],
\end{equation}
where $p$ is the Lagrange multiplier. For incompressible uniaxial tension,
\begin{equation} \label{eq:48}
\mathbf{F}=\mathrm{diag}\left( \lambda, \lambda^{-1/2},\lambda^{-1/2}\right), 
\quad 
\mathbf{C}=\mathbf{F}^\mathrm{T}\mathbf{F}=\mathrm{diag}\left( \lambda^2, \lambda^{-1},\lambda^{-1}\right).
\end{equation}

Imposing traction-free lateral conditions ($\mathbf{S}_{22}=\mathbf{S}_{33}=0$) yields an analytic expression for $p$ in terms of $\Gamma_1$ and $\Gamma_2$:
\begin{equation} \label{eq:49}
    p = \raisebox{2pt}{$\chi$}\left( W\right) \left(\frac{\Gamma_1(\bar{I}_1, \bar{I}_2)\left(\mathbb{G}_2\right)_{22} 
    + \Gamma_2(\bar{I}_1, \bar{I}_2)\left(\mathbb{G}_3\right)_{22}}{\left(\mathbb{G}_1\right)_{22}}\right).
\end{equation}
Substituting back into Eq.~(\ref{eq:47}) eliminates $p$ and leads to
\begin{equation} \label{eq:50}
\begin{aligned}
\mathbf{S} 
&= \raisebox{2pt}{$\chi$}\left( W\right) \mathbf{S}_\mathrm{intact}
\\ 
&= \raisebox{2pt}{$\chi$}\left( W\right)\Bigg[\Gamma_1(\bar{I}_1, \bar{I}_2)\left(\mathbb{G}_2 - \frac{\left(\mathbb{G}_2\right)_{22}}{\left(\mathbb{G}_1\right)_{22}} \mathbb{G}_1\right)
+ \Gamma_2(\bar{I}_1, \bar{I}_2)\left(\mathbb{G}_3 - \frac{\left(\mathbb{G}_3\right)_{22}}{\left(\mathbb{G}_1\right)_{22}} \mathbb{G}_1\right)\Bigg].
\end{aligned}
\end{equation}

In Stage I, the intact portion of the axial response is isolated from the experimental stress--stretch pairs using a user-defined stretch cutoff. Using Eq.~(\ref{eq:50}), the isochoric response functions are extracted via linear least squares at each data point by solving
\begin{equation} \label{eq:51}
\underbrace{
\begin{bmatrix}
\mathrm{vec}\left(\mathbb{G}_2 - \frac{\left(\mathbb{G}_2\right)_{22}}{\left(\mathbb{G}_1\right)_{22}} \mathbb{G}_1\right) & 
\mathrm{vec}\left(\mathbb{G}_3 - \frac{\left(\mathbb{G}_3\right)_{22}}{\left(\mathbb{G}_1\right)_{22}} \mathbb{G}_1\right)
\end{bmatrix}}_{A}
\underbrace{
\begin{bmatrix}
\Gamma_1 \\
\Gamma_2
\end{bmatrix}}_{x}
=
\underbrace{\mathrm{vec}(\mathbf{S}_{\mathrm{intact}})}_{b}.
\end{equation}

The isochoric GPR model $\mathcal{M}_{iso}:(\bar{I}_1,\bar{I}_2)\mapsto(\Gamma_1,\Gamma_2)$ is trained using the intact dataset. The intact strain energy density is then computed over the full stretch history (cf. Eq.~(\ref{eq:27})) as
\begin{equation}\label{eq:52}
W = \int_{1}^{\lambda}\left(\widetilde{\mathbf{S}}_\mathrm{intact}\right)_{11}\,d\lambda,
\end{equation}
where $\widetilde{(\cdot)}$ denotes predictions from the Stage I model. The damage factor $\raisebox{2pt}{$\chi$}(W)$ is subsequently obtained from the proportionality relation between measured and intact stresses (Eq.~(\ref{eq:29})), and Stage II trains the constrained GPR model $\mathcal{M}_{dam}:W \mapsto \raisebox{2pt}{$\chi$} $ following the general procedure described in Section~\ref{subsec:stage II model}. 

The predictive performance of the proposed two-stage physics-informed GPR-based model is evaluated by directly comparing model predictions with the experimental stress–stretch data using the relative error metric defined in Section~\ref{sec:4.3}. In contrast to the synthetic data validation study presented earlier, which provides systematic comparative benchmarking against other models across a range of deformation modes, the objective of this section is not comparative benchmarking but rather to demonstrate the practical applicability of the proposed framework to real, noisy, and biologically heterogeneous soft-tissue data. The model is trained and evaluated using only the available uniaxial tension measurements, and the resulting predictions are interpreted in terms of both fitting accuracy and the physical relevance and interpretability of the inferred elastic–damage decomposition.

\subsection{Model Evaluation Metric}
\label{sec:4.3}

To quantitatively assess model performance, the percentage relative error at each data point is computed using the Frobenius norm of the predicted and reference second Piola--Kirchhoff stress tensors:
\begin{equation} \label{eq:54}
e_i = 100 \times \frac{\|\widetilde{\mathbf{S}}_i - \mathbf{S}_i^{\mathrm{true}}\|_F}{\|\mathbf{S}_i^{\mathrm{true}}\|_F},
\end{equation}
where $\|\cdot\|_F$ denotes the Frobenius norm, $\widetilde{\mathbf{S}}_i$ is the model prediction, and $\mathbf{S}_i^{\mathrm{true}}$ is the reference stress tensor (synthetic ground truth or experimentally derived stress). For synthetic datasets, errors are reported separately for the training deformation mode (uniaxial tension) and the out-of-sample deformation modes (uniaxial compression and simple shear). For experimental datasets, evaluation is restricted to the available uniaxial tension regime.

\section{Results and Discussion}
\label{sec:results_and_discussion}

\subsection{Model Performance with Synthetic Mechanical Test Data}
\subsubsection{In-distribution accuracy: Uniaxial tension}
\label{subsubsec:5.1.1}

We first examine the in-distribution predictive performance of the proposed two-stage physics-informed GPR-based model using the synthetic uniaxial tension dataset described in Section~\ref{subsec:4.1}. Since the underlying constitutive response is known exactly, this benchmark enables a rigorous comparison between model predictions and the ground-truth behavior in both the intact and damage-dominated regimes.

\paragraph{\textbf{Training setup and implementation choices}}
The synthetic training dataset consists of $N=51$ deformation states uniformly sampled over the stretch range $\lambda \in [1 , 1.5]$. As described in Section~\ref{subsec:stage I model}, the intact (undamaged) portion of the response required for Stage~I training is isolated using a stretch cutoff of $\lambda = 1.25$. This choice ensures that the volumetric and isochoric response functions are learned exclusively from the undamaged regime, while still spanning a sufficiently rich range of deformation states.

Stage~II damage learning follows the constrained GPR framework detailed in Section~\ref{subsec:stage II model}. Monotonicity and non-negativity constraints on the stress-reduction factor $\raisebox{2pt}{$\chi$}(W)$ are enforced at a set of $N_c = 30$ constraint points selected over the strain-energy range $W \in [0.6, 1.25]$ (see Eq. (\ref{eq:29.5c})), corresponding approximately to $0.8W_{\mathrm{peak}} \leq W \leq 1.7W_{\mathrm{peak}}$. This interval spans the region where $\raisebox{2pt}{$\chi$}(W)$ transitions rapidly from unity toward zero and is therefore critical for stabilizing the damage evolution. Without these constraints, the inferred $\raisebox{2pt}{$\chi$}(W)$ may exhibit non-physical behavior, such as becoming negative or locally increasing before approaching zero. The role of these constraints is further examined in Section~\ref{subsubsec:5.1.4}.

Both the non-negativity and monotonicity penalties use weights of $\Lambda_{\mathrm{nn}}=\Lambda_{\mathrm{mono}}=10^3$ (Eqs.~(\ref{eq:29.5d}) and (\ref{eq:34})). Complete failure at large intact strain energy densities is enforced via data augmentation by introducing artificial points with $\raisebox{2pt}{$\chi$}=0$ in the range $W \in [1.3W_{\mathrm{peak}}, 2.6W_{\mathrm{peak}}]$ (i.e., $a=1.3$ and $b=2.6$ in the notation of Section~\ref{subsubsec:3.3.2}).

Finally, the nugget parameter $\alpha$ (Eq.~(\ref{eq:23})) is chosen separately for each GPR model to balance numerical conditioning and fidelity. For the volumetric model $\mathcal{M}_{vol}$, $\alpha = 10^{-5}$ is assigned at the stress-free reference state, while $\alpha = 10^{-2}$ is used for the remaining training points. For the isochoric model $\mathcal{M}_{iso}$, a larger nugget $\alpha = 10^{0}$ is used at all training points to enhance numerical stability and mitigate overfitting in the two-output regression. For the damage model $\mathcal{M}_{dam}$, $\alpha = 10^{-4}$ is used at all training points.

For consistency across model comparisons, nugget parameters are also specified for the baseline data-driven models. The black-box GPR model $\mathcal{M}_{black\text{-}box}$ employs a uniform nugget value of $\alpha = 10^{-2}$ at all training points. For the single-stage (direct) GPR-based model, the volumetric and isochoric mappings, $\mathcal{M}_{direct,vol}$ and $\mathcal{M}_{direct,iso}$, use the same nugget values as their corresponding Stage~I counterparts in the proposed framework, i.e., $\alpha = 10^{-5}$ at the stress-free reference state and $\alpha = 10^{-2}$ elsewhere for $\mathcal{M}_{direct,vol}$, and $\alpha = 10^{0}$ at all training points for $\mathcal{M}_{direct,iso}$. This choice ensures a fair and consistent comparison of model performance.

\begin{figure}[t]
    \centering
    \includegraphics[width=5in]{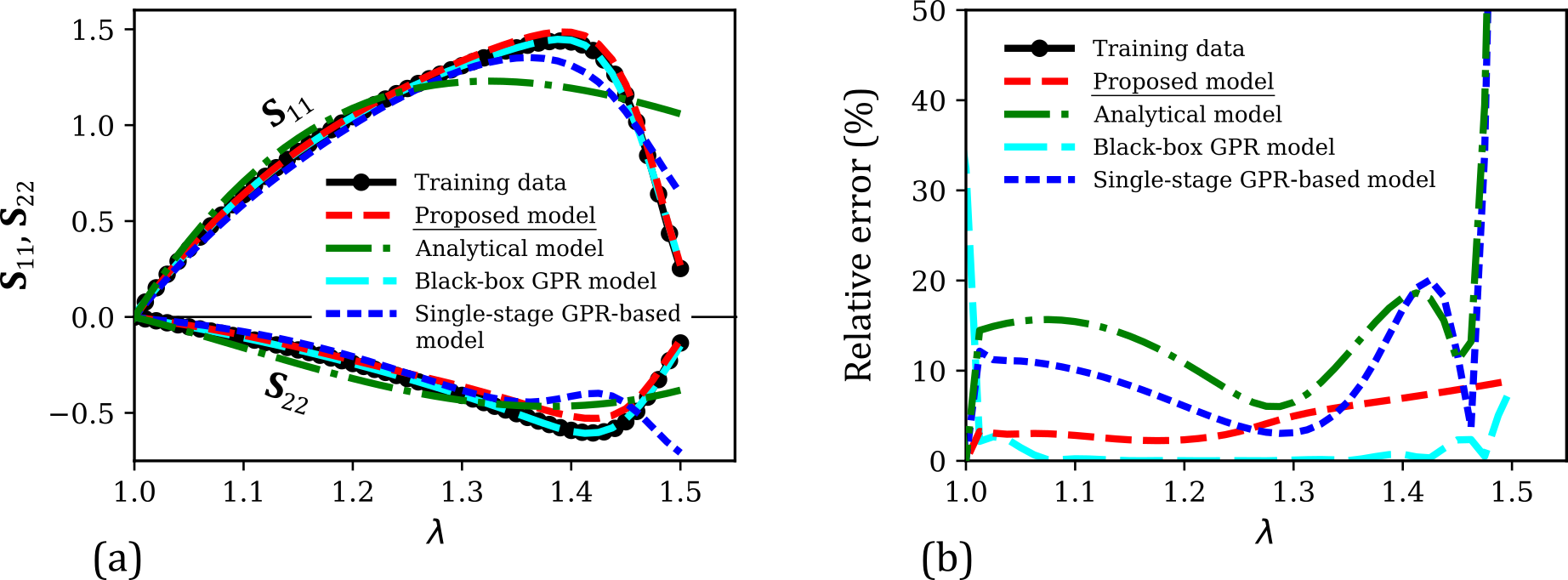}
    \caption{%
    In-distribution model performance under uniaxial tension using synthetic data. (a) Axial and lateral second Piola--Kirchhoff stress components, $\mathbf{S}_{11}$ and $\mathbf{S}_{22}$, versus stretch $\lambda$ in the training regime, comparing the proposed two-stage physics-informed GPR-based model with the analytical constitutive model, a black-box GPR model, and a single-stage (direct) GPR-based model. (b) Corresponding percentage relative error versus stretch, computed using the Frobenius norm of the stress tensor.
    }
    \label{fig:synthetic_fitting-results}
\end{figure}

\paragraph{\textbf{Stress–stretch predictions and error analysis}}
Figures~\ref{fig:synthetic_fitting-results}(a--b) jointly compare the predicted axial and lateral second Piola--Kirchhoff stress components and the corresponding pointwise percentage relative errors (Eq.~(\ref{eq:54})) as functions of stretch within the training regime for four models: (i) the proposed two-stage physics-informed GPR-based model, (ii) the analytical constitutive model in Eq.~(\ref{eq:43}), (iii) the black-box GPR model mapping in Eq.~(\ref{eq:44}), and (iv) the single-stage (direct) GPR-based model in Eqs.~(\ref{eq:45}) and (\ref{eq:46}), which maps invariants directly to scaled response functions.

In the intact regime ($\lambda \lesssim 1.25$), all data-driven models reproduce the ground-truth stress--stretch response with small errors, as evidenced by the close overlap of curves in Fig.~\ref{fig:synthetic_fitting-results}(a) and the uniformly low relative errors in Fig.~\ref{fig:synthetic_fitting-results}(b). As the response transitions into the damage-dominated regime, clear distinctions emerge. The black-box GPR model continues to interpolate the stress--stretch data almost exactly across the entire stretch range, resulting in the smallest pointwise errors overall. This behavior is reflected in Fig.~\ref{fig:synthetic_fitting-results}(b), where the black-box model consistently exhibits near-zero relative error except in the immediate vicinity of the reference state. This high fitting accuracy is expected, as the black-box model directly maps deformation tensor components to stress components without intermediate representations such as response functions or damage variables. Under uniaxial tension, the effective dimensionality of both the input and output spaces is low (two unique deformation and stress components), and thus the advantages of dimensionality reduction and physics-based decomposition in the proposed framework do not translate into superior interpolation accuracy in this specific in-distribution setting.

By contrast, the proposed two-stage model introduces an intermediate decomposition into intact hyperelastic response and damage evolution. While this structure slightly limits pure interpolation accuracy relative to the black-box approach, it yields a physically interpretable and thermodynamically admissible representation of the material response. As seen in Fig.~\ref{fig:synthetic_fitting-results}(a), this manifests as small deviations from the ground-truth stress response, which are mirrored by modest increases in relative error in Fig.~\ref{fig:synthetic_fitting-results}(b). Nevertheless, the proposed model captures both the nonlinear elasticity and softening behavior with high fidelity.

Importantly, the proposed model substantially outperforms the analytical constitutive model, whose fixed functional form limits its ability to capture the nonlinear transition from intact behavior to progressive softening. The single-stage GPR-based model also exhibits noticeable deviations near the peak and post-peak regions, where the response becomes highly nonlinear and damage-dominated, and thus, a direct mapping struggles to represent the evolving constitutive behavior. These discrepancies are evident in Fig.~\ref{fig:synthetic_fitting-results}(a) as underprediction or over-smoothing of the stress response and are directly reflected in the large relative error spikes in Fig.~\ref{fig:synthetic_fitting-results}(b).
\begin{table}[b]
\centering
\caption{Mean ($\bar{e}$) and maximum ($e_{max}$) percentage relative errors in the uniaxial tension training regime for the synthetic dataset, comparing the proposed two-stage physics-informed GPR-based model with an analytical constitutive model, a black-box GPR model, and the single-stage (direct) GPR-based model. Errors are computed using the relative metric defined in Eq.~(\ref{eq:54}).}
\label{tab:synthetic_training_errors}
\begin{tabular}{lcc}
\toprule
\small{\textbf{Model}} & \small{Mean error, $\bar{e}$ (\%)} & \small{Maximum error, $e_{max}$ (\%)}\\
\midrule
\small{Two-stage physics-informed GPR-based model (this work)}   & \small{4.52} & \small{8.92} \\
\small{Analytical constitutive model (Eq.~(\ref{eq:43}))}  & \small{21.65} & \small{276.71} \\
\small{Black-box GPR model (Eq.~(\ref{eq:44}))} & \small{1.62} & \small{32.44} \\
\small{Single-stage (direct) GPR-based model (Eq.~(\ref{eq:45}))} & \small{18.20} & \small{283.04} \\
\bottomrule
\end{tabular}
\end{table}

Table~\ref{tab:synthetic_training_errors} summarizes the mean and maximum percentage relative errors over the training regime. While the black-box GPR model yields the smallest mean error, its comparatively high maximum error is not indicative of poor performance in the damage regime; rather, it is concentrated near the stress-free reference configuration, where the denominator in the relative error metric becomes very small and even minor absolute deviations are amplified. Overall, the proposed two-stage model maintains consistently low errors across both intact and damage-dominated portions of the response, substantially outperforming the analytical and single-stage GPR-based models while retaining a physically interpretable structure. Although it does not surpass the black-box GPR in pure interpolation accuracy under uniaxial tension, it achieves a more favorable balance between flexibility and physical structure. As shown next in Section~\ref{subsubsec:5.1.3}, this balance is critical for achieving robust and physically meaningful predictions in out-of-sample deformation modes, where purely data-driven black-box models fail to generalize.

\subsubsection{Process-level interpretation of the proposed two-stage model}

\begin{figure}[t]
    \centering
    \includegraphics[width=6.3in]{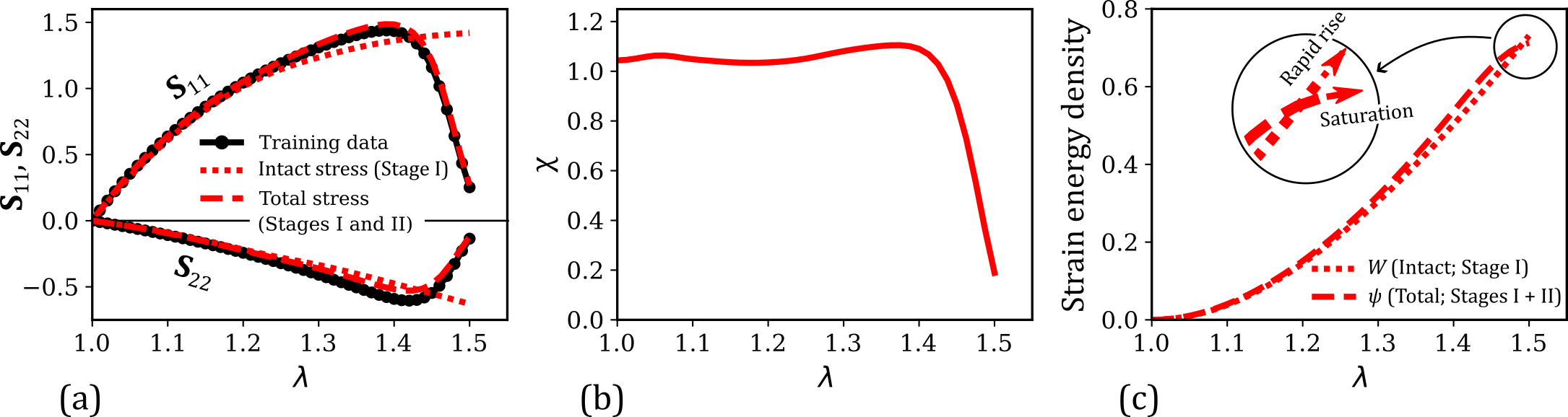}
    \caption{%
    Process-level interpretation of the two-stage physics-informed GPR-based constitutive model under uniaxial tension (synthetic data).
    (a) Axial (11) and lateral (22) components of the predicted intact stress $\widetilde{\mathbf{S}}_{\mathrm{intact}}$ from the Stage I GPR models and the total stress $\widetilde{\mathbf{S}}$ from the full two-stage model, compared with training data.
    (b) Stress-reduction factor $\raisebox{2pt}{$\chi$} $ predicted by the proposed two-stage model as a function of stretch $\lambda$, showing monotonic, non-negative decay toward zero.
    (c) Predicted intact hyperelastic strain energy density $W$ (from Stage I GPR models) and total strain energy density $\psi$ (from the complete model) versus stretch, illustrating energy saturation in the damage regime while the intact energy continues to increase.
    }
    \label{fig:synthetic_process-level}
\end{figure}

Beyond predictive accuracy, a key objective of the proposed framework is to provide a physically interpretable description of material behavior by explicitly separating elastic response from progressive damage. Figure~\ref{fig:synthetic_process-level} illustrates this process-level decomposition for the synthetic uniaxial tension dataset and highlights how the two-stage physics-informed GPR-based model recovers meaningful internal variables that are absent in competing data-driven approaches.

\paragraph{\textbf{Separation of intact elasticity and damage effects}}
Figure~\ref{fig:synthetic_process-level}(a) compares the axial and lateral components of the predicted intact (undamaged) second Piola--Kirchhoff stress $\widetilde{\mathbf{S}}_{\mathrm{intact}}$ obtained from the trained Stage~I GPR models $\mathcal{M}_{vol}$ and $\mathcal{M}_{iso}$ (Eq. (\ref{eq:26.5})) with the total stress produced by the full two-stage model (Eq. (\ref{eq:18})). In the intact regime ($\lambda \lesssim 1.25$), the intact stress closely follows the training data, confirming that the Stage~I GPR models accurately learn the underlying hyperelastic response. As deformation proceeds into the damage-dominated regime, the intact stress continues to increase smoothly, reflecting the extrapolated elastic response that would be obtained in the absence of damage. The final stress response, by contrast, deviates from this intact trajectory and exhibits progressive softening and eventual failure due to the multiplicative action of the stress-reduction factor learned in Stage~II.

This clear divergence between intact and total stress responses demonstrates that the proposed framework achieves a clean and explicit separation between elasticity and damage. Importantly, this separation is not imposed through a prescribed analytical form but is learned directly from data while respecting fundamental physical constraints.

\paragraph{\textbf{Damage evolution through a physically meaningful internal variable}}
The evolution of the stress-reduction factor $\raisebox{2pt}{$\chi$} $ predicted by the trained Stage~II constrained GPR model $\mathcal{M}_{dam}$ is shown in Fig.~\ref{fig:synthetic_process-level}(b). The inferred $\raisebox{2pt}{$\chi$} $ versus $\lambda$ response exhibits a monotonic, non-negative decay from values close to unity toward zero as deformation increases. This behavior is consistent with the physical interpretation of $\raisebox{2pt}{$\chi$} $ as a scalar measure of material integrity. Although $\raisebox{2pt}{$\chi$} $ is not exactly equal to unity in the intact regime, owing to small prediction errors in the Stage~I GPR models used to compute $\widetilde{\mathbf{S}}_{\mathrm{intact}}$, it remains within approximately 10\% of unity prior to damage onset (i.e., prior to $\lambda\approx1.4$, which corresponds to the peak axial stress). Crucially, $\raisebox{2pt}{$\chi$} $ never becomes negative, nor does it exhibit any increase in the post-peak region, which would indicate non-physical material healing.

The smooth decay of $\raisebox{2pt}{$\chi$} $ toward zero reflects progressive damage accumulation and eventual failure, without spurious oscillations or reversals. As discussed briefly in Section~\ref{subsubsec:5.1.1} and examined in detail in Section~\ref{subsubsec:5.1.4}, this physically admissible evolution is a direct consequence of the monotonicity and non-negativity constraints imposed during Stage~II training. Without these constraints, the learned damage evolution can become non-physical, highlighting the necessity of the physics-informed penalty formulation.

\paragraph{\textbf{Energy-based interpretation of damage and failure}}
Figure~\ref{fig:synthetic_process-level}(c) provides an energy-based perspective on the two-stage decomposition by comparing the predicted intact hyperelastic strain energy density $W$ (Eq.~(\ref{eq:27})) with the total strain energy density $\psi$, computed using the same integral definition as Eq.~(\ref{eq:27}) but with the intact stress $\widetilde{\mathbf{S}}_{\mathrm{intact}}$ replaced by the total predicted stress $\widetilde{\mathbf{S}}$ from the full two-stage model. In the intact regime, the two energy measures are nearly identical and both exhibit convex growth with stretch, indicating that damage effects are negligible and the response is dominated by elastic energy storage. As deformation increases beyond the onset of damage, the two curves begin to diverge. The intact strain energy density continues to increase in a convex manner, reflecting the underlying hyperelastic response. In contrast, the total strain energy density saturates and approaches an asymptotic limit, as clearly illustrated in the inset of Fig.~\ref{fig:synthetic_process-level}(c).

This saturation behavior is a hallmark of energy-limited hyperelastic models and provides a physically meaningful description of material failure: additional mechanical work beyond a critical threshold is dissipated into damage rather than stored elastically. The fact that the proposed two-stage physics-informed GPR-based model reproduces this behavior directly from data—without prescribing an explicit analytical damage law—underscores its ability to discover physically consistent constitutive structure.

\paragraph{\textbf{Comparison with competing data-driven models}}
The process-level interpretability demonstrated in Fig.~\ref{fig:synthetic_process-level} is absent in competing data-driven approaches. Black-box GPR models provide no explicit internal damage variable and therefore cannot distinguish between elastic response and damage-induced softening. Similarly, the single-stage (direct) GPR-based model entangles elasticity and damage within a single mapping, making it impossible to isolate damage onset, progression, or failure in a physically meaningful way. In contrast, the proposed two-stage model introduces a clear and interpretable internal variable, $\raisebox{2pt}{$\chi$} $, which governs damage evolution in a manner consistent with the energy-limited hyperelasticity framework.

In this sense, the proposed framework combines the strengths of both analytical and data-driven constitutive modeling. Like classical hyperelasticity models with energy limiters, it provides a structured, interpretable description of elasticity and damage. At the same time, it retains the flexibility, automation, and high fitting accuracy of ML-based approaches, without requiring expert intervention or manual model selection. As shown next in Section~\ref{subsubsec:5.1.3}, these advantages extend beyond the training regime and enable robust, physically meaningful generalization to deformation modes not included during training, further reinforcing the ability of the proposed model to unify interpretability and generalizability within a single data-driven constitutive framework.

\subsubsection{Out-of-sample generalization: Compression and shear}
\label{subsubsec:5.1.3}

We next assess the out-of-sample generalization capability of the proposed two-stage physics-informed GPR-based model by evaluating its predictions under deformation modes not used during training, namely uniaxial compression and simple shear (Eq. (\ref{eq:42})). The purpose of this study is not to demonstrate numerical accuracy per se, but to examine whether the learned constitutive structure extrapolates in a physically meaningful manner beyond the calibration regime. All reference responses shown in this section are generated using the same underlying constitutive model employed for synthetic data generation (i.e., Eq.~(\ref{eq:41})) and are therefore treated as ground truth.

\begin{figure}[t]
    \centering
    \includegraphics[width=6.3in]{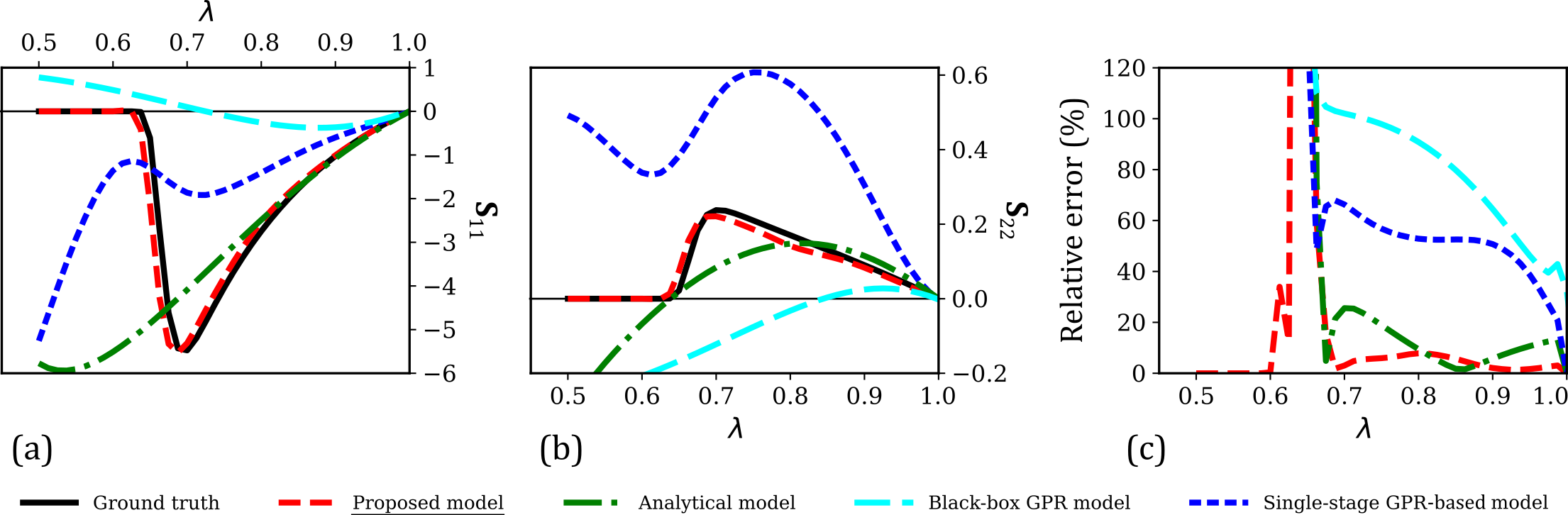}
    \caption{%
    Out-of-sample generalization under uniaxial compression using synthetic data. (a) Axial second Piola–Kirchhoff stress component $\mathbf{S}_{11}$ and (b) lateral stress component $\mathbf{S}_{22}$ versus stretch $\lambda \in [1,0.5]$, comparing predictions of the proposed two-stage physics-informed GPR-based model with the analytical constitutive model, black-box GPR model, and single-stage (direct) GPR-based model against the ground-truth response. (c) Corresponding pointwise percentage relative error as a function of stretch, computed using the Frobenius norm of the stress tensor.
    }
    \label{fig:synthetic_compression}
\end{figure}

\paragraph{\textbf{Performance in uniaxial compression}}
Figures~\ref{fig:synthetic_compression}(a--b) show the predicted axial and lateral second Piola--Kirchhoff stress components, $\mathbf{S}_{11}$ and $\mathbf{S}_{22}$, for all four models under comparison, as functions of stretch $\lambda \in [1,0.5]$ in uniaxial compression. Figure~\ref{fig:synthetic_compression}(c) reports the corresponding pointwise percentage relative errors. The ground-truth response exhibits an initially nonlinear elastic regime from $\lambda=1$ down to approximately $\lambda \approx 0.7$, followed by progressive softening and complete stress collapse near $\lambda \approx 0.64$, reflecting compressive failure. Notably, the elastic portion of the response is stiffer in compression than in tension (cf. Fig. \ref{fig:synthetic_fitting-results}), illustrating tension–compression asymmetry, a common feature of many soft materials arising from microstructural constraints and volumetric resistance under compressive loading \cite{Du2020}.

The proposed model accurately reproduces this qualitative behavior. As seen in Figs.~\ref{fig:synthetic_compression}(a--b), both axial and lateral stress components rise nonlinearly in compression, followed by a smooth decay to zero stress, after which the stresses remain identically zero. This behavior is consistent with the expected evolution of a damage-governed material and reflects the combined effect of the intact elastic response learned in Stage~I and the monotonic, non-negative stress-reduction factor learned in Stage~II. The model also captures the tension–compression asymmetry observed in the elastic regime, despite being trained exclusively on tensile data. Notably, these predictions are obtained using only $N=51$ training data points from uniaxial tension.

The analytical constitutive model also predicts a rising-and-falling response under compression; however, its lateral stress component $\mathbf{S}_{22}$ changes sign near $\lambda \approx 0.64$. Although the magnitude of $\mathbf{S}_{22}$ is small relative to $\mathbf{S}_{11}$, this sign reversal is physically questionable and indicates a limitation of the fixed functional form in capturing multiaxial damage effects.

The single-stage (direct) GPR-based model exhibits more severe deficiencies. While its predicted stresses initially rise and begin to soften, the response subsequently stiffens again at larger compressive strains, leading to a non-monotonic stress evolution. This behavior is unphysical and directly reflects the absence of an explicit damage variable and monotonicity constraints in the single-stage mapping.

The black-box GPR model performs the poorest in compression. Both $\mathbf{S}_{11}$ and $\mathbf{S}_{22}$ predicted by this model change sign, and most notably, the model predicts tensile axial stress under compressive deformation. Furthermore, no failure is predicted, and stresses remain nonzero at large compressive strains. These results highlight the fundamental limitation of purely data-driven mappings when extrapolated beyond the training regime.

The percentage relative error plot in Fig.~\ref{fig:synthetic_compression}(c) confirms these trends: although absolute error magnitudes are not emphasized here—given that the model was calibrated using tensile data only—the proposed model consistently exhibits the lowest errors among all models. Error spikes near the failure point are observed for all models and are expected, as the denominator in the relative error metric approaches zero when stresses vanish (see Eq.~(\ref{eq:54})).

Although strain energy density evolution is not shown explicitly in this section, the stress–stretch responses in Figs.~\ref{fig:synthetic_compression}(a--b) implicitly reveal important energetic differences between the models. Both the proposed model and the analytical constitutive model exhibit clear saturation of the total strain energy density as deformation progresses, consistent with their explicit energy-limiting formulations. In contrast, neither the black-box GPR model nor the single-stage GPR model enforces such energetic bounds, leading to non-physical stress persistence or re-stiffening at large compressive strains. Importantly, the proposed model is the only data-driven approach that simultaneously reproduces the correct stress trends and the expected energy-limited failure behavior.

\begin{figure}[t]
    \centering
    \includegraphics[width=5in]{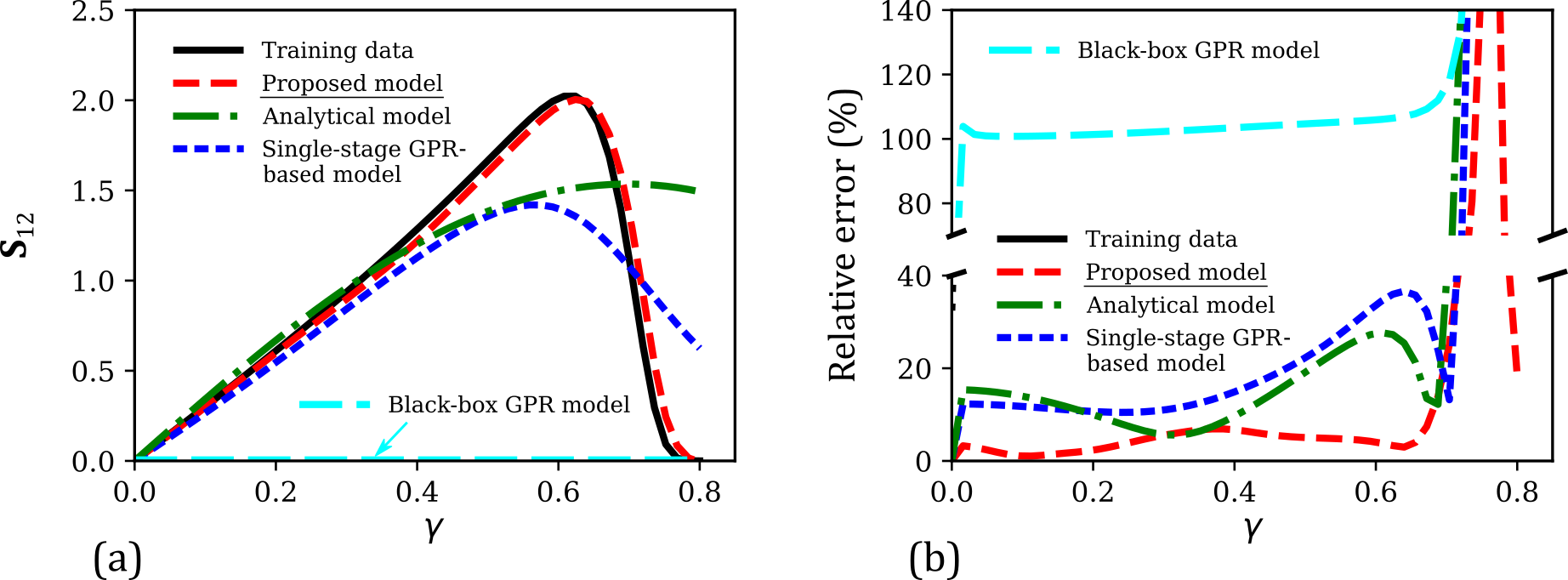}
    \caption{%
    Out-of-sample generalization under simple shear using synthetic data. (a) Predicted shear stress component $\mathbf{S}_{12}$ versus shear strain $\gamma$, and (b) corresponding pointwise percentage relative error versus $\gamma$, comparing the proposed two-stage physics-informed GPR-based model with the analytical constitutive model, black-box GPR model, and single-stage (direct) GPR-based model against the ground-truth response.
    }
    \label{fig:synthetic_shear}
\end{figure}

\paragraph{\textbf{Performance in simple shear}}
Figure~\ref{fig:synthetic_shear}(a) shows the predicted shear stress component $\mathbf{S}_{12}$ for all four models under comparison as a function of shear strain $\gamma$, while Fig.~\ref{fig:synthetic_shear}(b) reports the corresponding percentage relative error. Simple shear represents a particularly stringent generalization test, as it involves off-diagonal components of the deformation gradient that are entirely absent from the uniaxial tension training data ($N=51$ data points).

Despite this, the proposed model accurately captures the ground-truth response. The predicted shear stress exhibits a nonlinear elastic rise followed by softening and eventual stress collapse, in close agreement with the reference solution. This behavior translates to low relative errors across most of the deformation range, with elevated errors confined to the post-peak regime where stresses decrease rapidly. As in compression, this stress evolution implies a bounded and saturating total strain energy density under shear for the proposed model, whereas such energetic consistency is absent in the black-box GPR and single-stage GPR-based models.

The analytical constitutive model and the single-stage GPR-based model both predict a qualitatively reasonable rise-and-fall response under shear, indicating physically admissible elastic loading followed by softening and failure. While their peak magnitudes and softening rates differ from the ground-truth response, these deviations are less critical here, as the objective of this investigation is not quantitative accuracy but physical plausibility under unseen deformation modes.

In contrast, the black-box GPR model exhibits a striking failure mode: it predicts $\mathbf{S}_{12}=0$ for all values of $\gamma$. This result is a direct consequence of its construction. Because the model is trained exclusively on uniaxial tension data, where both the deformation tensor $\mathbf{C}$ and the stress tensor $\mathbf{S}$ are diagonal, it has no information about how off-diagonal components should influence the stress response. As a result, the model defaults to zero shear stress, yielding extremely large relative errors in Fig.~\ref{fig:synthetic_shear}(b). Notably, this same model was the most accurate interpolator in the training regime (Section~\ref{subsubsec:5.1.1}), underscoring the disconnect between interpolation accuracy and physical generalizability.

\paragraph{\textbf{Assessment of out-of-sample model performance}}
Taken together, the compression and shear results demonstrate that the proposed two-stage physics-informed GPR-based model generalizes robustly to deformation modes not used for training. Unlike black-box models, which learn a high-dimensional but fragile mapping between tensor components, the proposed framework operates in a reduced invariant space and reconstructs stress through a physically motivated tensor basis. The explicit two-stage decomposition further enables direct enforcement of physically admissible damage evolution through monotonicity, non-negativity, and complete-failure constraints.

As a result, the proposed model reproduces correct qualitative trends—elastic loading, damage onset, softening, and failure—across both compression and shear, even though these deformation modes are entirely absent from the training data. Importantly, this robust generalization is achieved using a very small training dataset ($N=51$), demonstrating that the proposed model is highly data-efficient despite being ML-based, and is therefore well suited for solid mechanics and materials applications where large, diverse datasets are rarely available. The next subsection, Section~\ref{subsubsec:5.1.4}, further examines the role of constrained optimization and penalty-based learning in achieving this robustness.
\begin{figure}[t]
    \centering
    \includegraphics[width=4.94in]{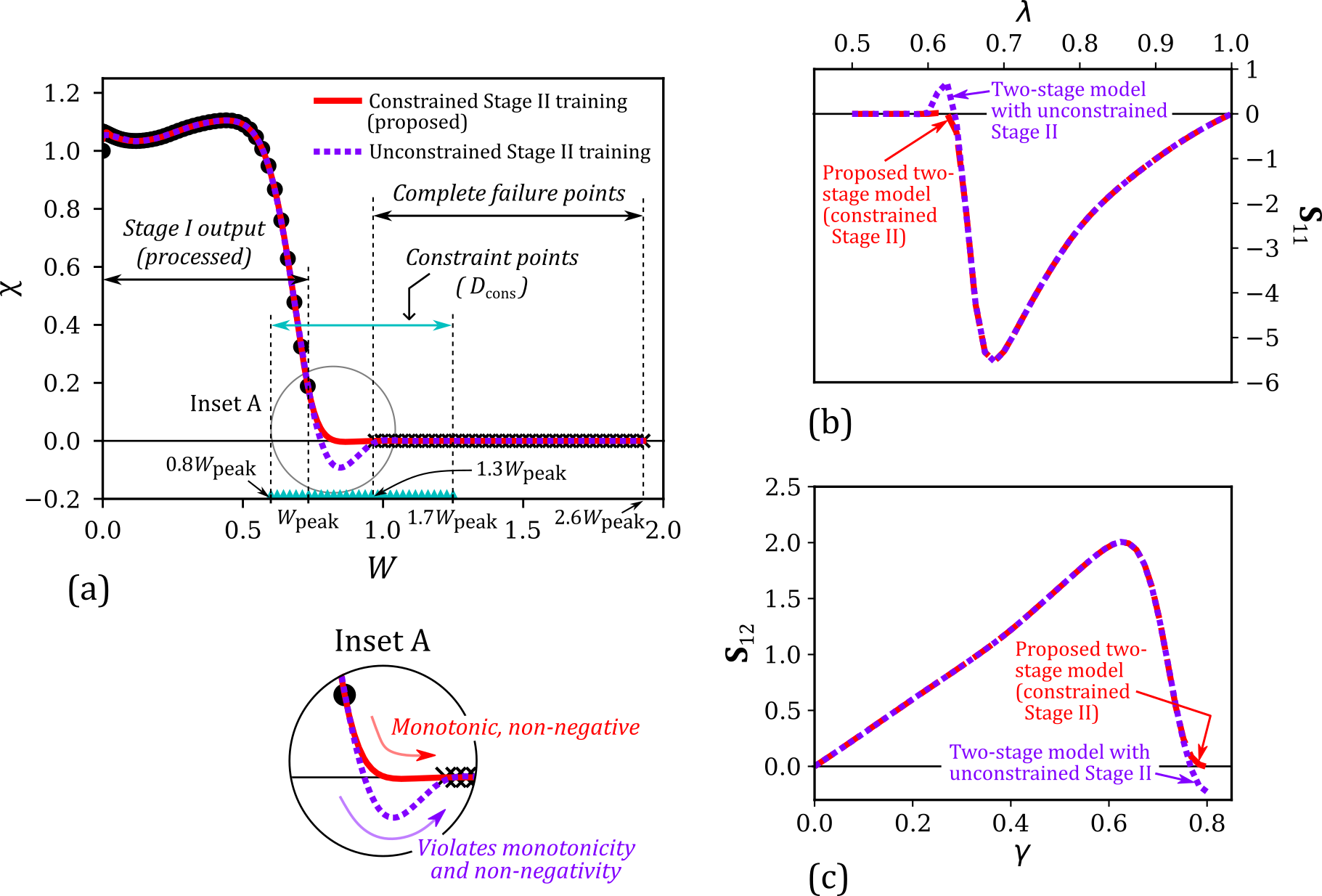}
    \caption{%
    Role of penalty-based constrained optimization during Stage~II damage learning. (a) Learned stress-reduction factor $\raisebox{2pt}{$\chi$}(W)$ as a function of intact hyperelastic strain energy density $W$, comparing the proposed framework with constrained Stage~II training against an otherwise identical framework trained without constraints. Black markers denote processed Stage~I outputs ($\mathcal{D}^*_{\mathrm{dam}}$ prior to augmentation), cyan markers indicate constraint points $\mathcal{D}_{\mathrm{cons}}$, and artificial complete-failure points ($\raisebox{2pt}{$\chi$}=0$) are added at large $W$. The constrained Stage~II training enforces monotonicity and non-negativity of $\raisebox{2pt}{$\chi$}(W)$, while unconstrained training leads to non-physical behavior (Inset~A). (b) Axial second Piola--Kirchhoff stress $\mathbf{S}_{11}$ versus compressive stretch $\lambda$, and (c) shear stress component $\mathbf{S}_{12}$ versus shear strain $\gamma$, comparing full two-stage model predictions obtained with constrained and unconstrained Stage~II training.
    }
    \label{fig:constrained_vs_unconstrained}
\end{figure}

\subsubsection{Role of penalty-based constrained optimization in Stage~II learning}
\label{subsubsec:5.1.4}
This subsection examines the role of penalty-based constrained optimization during Stage~II training of the damage model $\mathcal{M}_{\mathrm{dam}}$ (Section~\ref{subsec:stage II model}). In particular, we compare the proposed two-stage framework—where Stage~II training enforces monotonicity and non-negativity of the stress-reduction factor $\raisebox{2pt}{$\chi$}(W)$—with an otherwise identical variant in which Stage~II is trained without such constraints. This comparison isolates the effect of constraint enforcement on damage evolution and its consequences for stress prediction, both within and beyond the training regime.

\paragraph{\textbf{Effect of constraints on the learned damage law}}
Figure~\ref{fig:constrained_vs_unconstrained}(a) illustrates the learning process and resulting predictions of the Stage~II damage model. The black circular markers correspond to the processed Stage~I outputs used to construct the initial damage dataset $\mathcal{D}^{\ast}_{\mathrm{dam}}$ (Eq.~(\ref{eq:29.5a})), where the intact strain energy density $W$ is computed from intact stress predictions of Stage I models using Eq.~(\ref{eq:27}), and the corresponding $\raisebox{2pt}{$\chi$}$ values are obtained by solving Eq.~(\ref{eq:29}). The maximum value of $W$ in this dataset is denoted by $W_{\mathrm{peak}}$. Also shown are the artificial complete-failure points with $\raisebox{2pt}{$\chi$}=0$, introduced over the range $W \in [1.3W_{\mathrm{peak}},\,2.6W_{\mathrm{peak}}]$ to enforce asymptotic failure. Together, the processed Stage~I outputs and these artificial points form the augmented dataset used for Stage~II training.

As mentioned previously, monotonicity and non-negativity constraints are enforced during Stage~II training through penalty-based optimization (Eq.~(\ref{eq:35.5})) at a discrete set of constraint points $\mathcal{D}_{\mathrm{cons}}$ (Eq.~(\ref{eq:29.5c})), shown by cyan markers in Fig.~\ref{fig:constrained_vs_unconstrained}(a), spanning the range $W \in [0.8W_{\mathrm{peak}},\,1.3W_{\mathrm{peak}}]$. This interval corresponds to the critical damage transition region where $\raisebox{2pt}{$\chi$}(W)$ rapidly decreases from near unity toward zero.

The red solid curve in Fig.~\ref{fig:constrained_vs_unconstrained}(a) shows the prediction obtained using the constrained Stage~II model (proposed framework). The stress-reduction factor decreases monotonically, remains non-negative, and asymptotically approaches zero, representing progressive damage accumulation followed by complete failure. This behavior is highlighted in the zoomed Inset~A. In contrast, the purple dashed curve corresponds to the same Stage~II model trained without penalty-based constraints (i.e., standard GPR optimization as in Eq.~(\ref{eq:24})). Despite access to identical data, the unconstrained training produces a non-physical evolution in which $\raisebox{2pt}{$\chi$}(W)$ becomes negative and locally increases before eventually returning to zero. Such behavior violates thermodynamic consistency and cannot be prevented without explicit constraint enforcement.

\paragraph{\textbf{Impact on out-of-sample stress predictions}}
Figures~\ref{fig:constrained_vs_unconstrained}(b) and \ref{fig:constrained_vs_unconstrained}(c) show the corresponding stress predictions under uniaxial compression and simple shear, respectively, comparing the proposed framework (with constrained Stage~II training) against the otherwise identical framework with unconstrained Stage~II training. Importantly, neither compression nor shear data are used during training.

In both deformation modes, the two frameworks produce nearly identical stress responses throughout the elastic regime and early stages of softening. However, pronounced differences emerge as deformation approaches failure. The proposed framework predicts a smooth and monotonic decay of stress to zero, consistent with complete material failure. In contrast, unconstrained Stage~II training leads to stress sign reversals and partial stress recovery near failure, indicative of spurious healing behavior. These non-physical responses directly reflect the unconstrained evolution of $\raisebox{2pt}{$\chi$}(W)$ observed in Fig.~\ref{fig:constrained_vs_unconstrained}(a).

Crucially, although the monotonicity and non-negativity constraints are enforced using tensile data alone, their formulation in terms of the scalar intact strain energy density $W$ ensures that they remain active across all deformation modes. This highlights a key advantage of the proposed framework: damage evolution is governed by an invariant-based internal variable rather than deformation-path-specific quantities, enabling consistent extrapolation to unseen loading conditions.

\paragraph{\textbf{Choice of constraint range and implications}}
The results in Fig.~\ref{fig:constrained_vs_unconstrained} also justify the selected range of constraint points $\mathcal{D}_{\mathrm{cons}}$. By concentrating constraint enforcement in the vicinity of the damage transition region ($W \in [0.8W_{\mathrm{peak}},\,1.3W_{\mathrm{peak}}]$), the proposed approach ensures physically admissible behavior where it is most critical, while avoiding unnecessary over-constraining of the regression problem. Enforcing constraints over a much wider range could lead to increased computational cost, numerical ill-conditioning, or over-regularization without providing additional physical benefit.

Overall, these results demonstrate that penalty-based constrained optimization during Stage~II GPR training is essential for achieving thermodynamically consistent damage evolution and physically meaningful stress predictions in both training and out-of-sample regimes. The constrained formulation allows the proposed two-stage physics-informed GPR-based model to retain the flexibility of data-driven learning while enforcing fundamental physical principles, thereby playing a central role in the robustness and generalizability of the framework.

%
%

\subsection{Application to Brain Tissue Mechanics}

We next apply the proposed two-stage physics-informed GPR-based model to uniaxial tension data of brain tissue reported by Franceschini et al.~\cite{Franceschini2006} (Fig.~\ref{fig:training_framework}(b)). This application demonstrates the practical utility of the framework for real, highly nonlinear soft-tissue data and assesses its ability to extract physically meaningful elastic and damage characteristics from limited experimental measurements. As detailed in Section~\ref{sec:4.2}, owing to the lack of volumetric response data in this case (a common limitation in soft tissue mechanics), the Stage~I protocol is modified to explicitly enforce incompressibility: only the isochoric model $\mathcal{M}_{iso}$ is learned via GPR, while the volumetric contribution is replaced by a deterministic Lagrange multiplier enforcing $J=1$. Stage~II damage modeling and constrained GPR learning of the stress-reduction factor remain unchanged.

\paragraph{\textbf{Training setup and implementation choices}}
Four experimental datasets are considered: two gray matter (GM) specimens from the thalamus region, and two white matter specimens obtained from distinct brain substructures---the occipital (WM-o) and parietal (WM-p) lobes. Each experimental stress--stretch curve is interpolated to $N=101$ uniformly spaced deformation states. A higher resolution than that used in synthetic validation ($N=51$) is adopted here to better capture local features present in the measured responses.
\begin{figure}[t]
    \centering
    \includegraphics[width=6.3in]{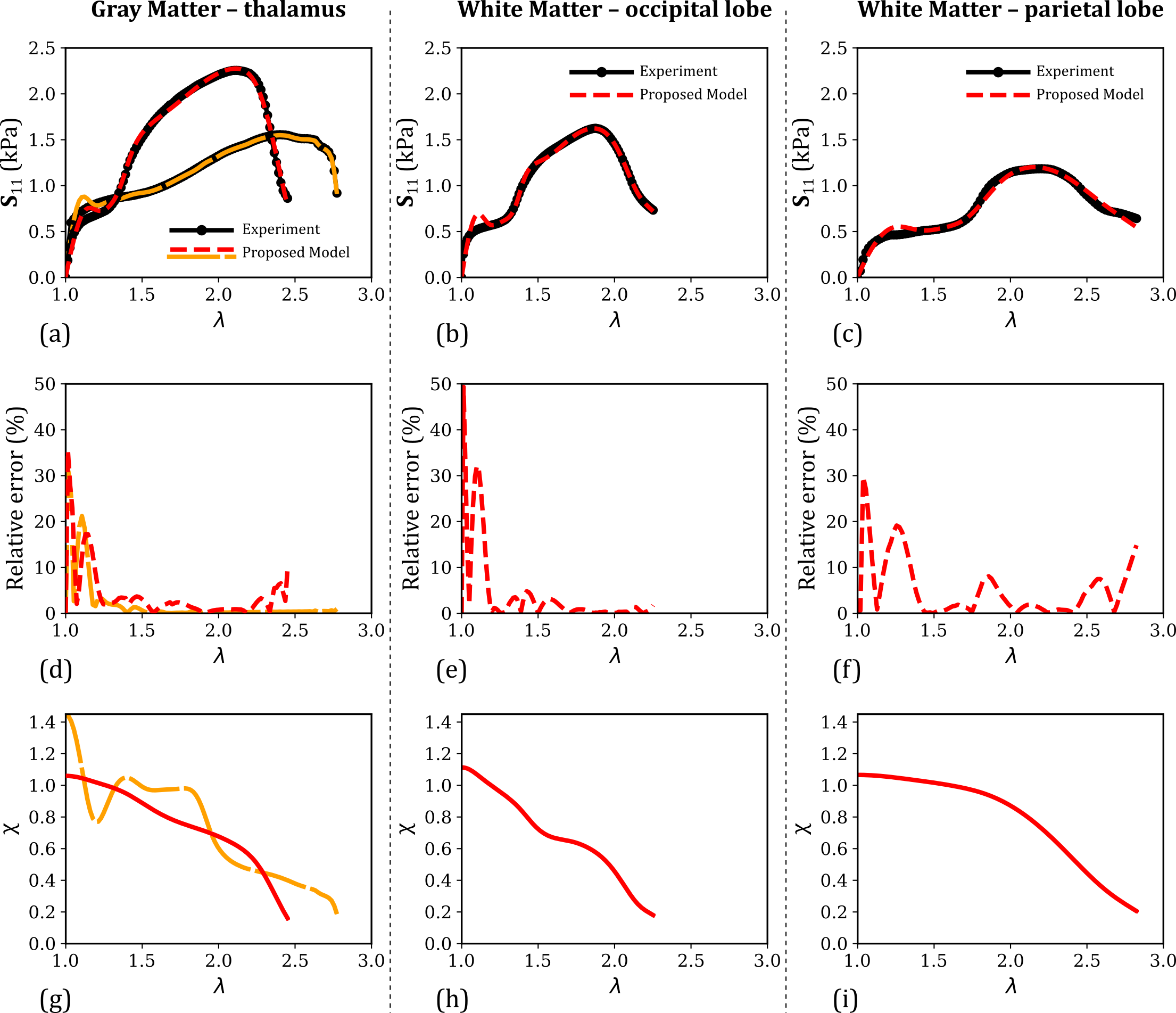}
    \caption{%
    Application of the proposed two-stage physics-informed GPR-based model to experimental uniaxial tension data of brain tissue (from \cite{Franceschini2006}). (a--c) Axial second Piola--Kirchhoff stress $\mathbf{S}_{11}$ versus stretch $\lambda$ for gray matter (GM) from the thalamus, white matter from the occipital lobe (WM-o), and white matter from the parietal lobe (WM-p), comparing experimental measurements with model predictions. (d--f) Corresponding pointwise percentage relative error as a function of stretch. (g--i) Predicted stress-reduction factor $\raisebox{2pt}{$\chi$}$ as a function of stretch for the three brain tissue substructures.    }
    \label{fig:experimental-evaluation}
\end{figure}

Owing to substantial variability in elastic stiffness and damage onset across these datasets, the Stage~I stretch cutoff used to isolate the intact response is selected individually for each case based on visual inspection of the stress--stretch curves. Specifically, the cutoffs are $\lambda = 1.89$ and $1.38$ for the two GM datasets, and $\lambda = 1.37$ and $1.80$ for WM-o and WM-p, respectively.

All remaining implementation choices are consistent across datasets. In Stage~II, monotonicity and non-negativity constraints on the stress-reduction factor are enforced at $N_c = 30$ constraint points spanning the strain-energy range $W \in [6,15]~\mathrm{kPa}$ (equivalently, kJ/m$^3$), with penalty weights $\Lambda_{\mathrm{nn}}=\Lambda_{\mathrm{mono}}=10^3$. Complete failure is imposed through data augmentation in the range $W \in [1.3W_{\mathrm{peak}}, 2.6W_{\mathrm{peak}}]$, mirroring the synthetic validation protocol.

For the GPR nugget parameters, $\alpha=0.5$ is used for the isochoric model $\mathcal{M}_{iso}$ at all training points, while $\alpha=2.5\times10^{-3}$ is used for the damage model $\mathcal{M}_{dam}$. These values are selected to balance fitting accuracy with numerical stability and to mitigate overfitting in the presence of highly nonlinear data.

\paragraph{\textbf{Stress–stretch fitting performance}}
Figure~\ref{fig:experimental-evaluation} evaluates the ability of the proposed model to reproduce experimental stress–stretch behavior across brain tissue substructures. The first row (Figs.~\ref{fig:experimental-evaluation}(a–c)) compares the experimentally measured axial stress $\mathbf{S}_{11}$ with the model predictions for GM, WM-o, and WM-p, while the second row (Figs.~\ref{fig:experimental-evaluation}(d–f)) reports the corresponding percentage relative error as a function of stretch.

Across all datasets, the proposed model reproduces the overall elastic response and progressive softening leading to failure with high fidelity. Nevertheless, a systematic deviation is observed in the early portion of the stress–stretch response. Specifically, the model overpredicts the experimental stress in the vicinity of a pronounced stress jump that appears across all datasets, where the response exhibits a very high initial stiffness followed by a rapid reduction in slope and a more gradual stress increase. This yield-like transition leads to a localized amplification of relative error in this region, as seen in Figs.~\ref{fig:experimental-evaluation}(d–f).

Although this behavior can be mitigated by selecting a smaller nugget parameter in Stage~I, doing so was found to induce overfitting in the intact regime and physically inadmissible behavior in the learned stress response near the intact stretch cutoff. In particular, excessively small nugget values lead to a spurious stress discontinuity at the cutoff stretch in the predicted intact response, which propagates into the total stress prediction and can lead to elevated errors and potential numerical issues in simulations (e.g., through the associated stiffness tensor). A sensitivity study illustrating this tradeoff between local interpolation accuracy in the intact regime and global physical robustness is presented in Supplementary Section~S1, where intact and total stress–stretch predictions are compared for different choices of the nugget parameter in the $\mathcal{M}_{iso}$ model using a representative brain tissue dataset.

For the nugget values adopted in the present study (with $\alpha=0.5$ for $\mathcal{M}_{iso}$), the mean percentage relative errors are low across all tissue types: 2.93\% (GM, averaged across two datasets), 4.06\% (WM-o), and 5.11\% (WM-p). The corresponding maximum errors—33.21\% (GM, averaged), 49.34\% (WM-o), and 29.77\% (WM-p)—are concentrated near the reference state and therefore correspond to small absolute stress discrepancies.
\begin{figure}[t]
    \centering
    \includegraphics[width=2.475in]{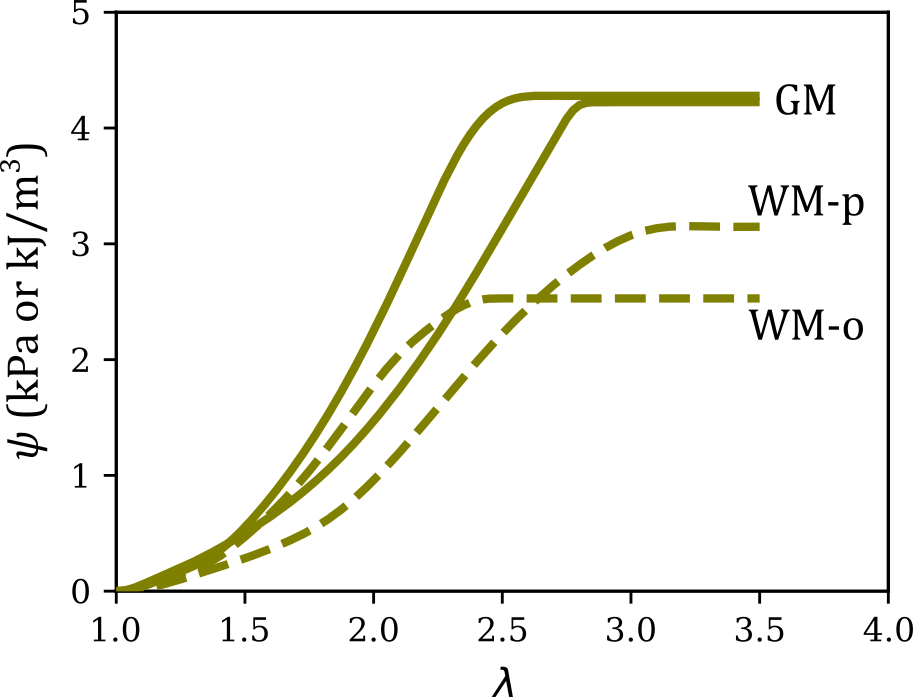}
    \caption{%
    Predicted total strain energy density $\psi$ versus stretch $\lambda$ for brain tissue under uniaxial tension, shown for two gray matter (GM) datasets and two white matter substructures (WM-o and WM-p). Predictions are obtained by extrapolating the trained two-stage physics-informed GPR-based model beyond the experimental range. In all cases, $\psi(\lambda)$ asymptotically approaches a saturation value, identifying the critical failure energy $\psi_f$ associated with complete material failure due to energy-limited damage evolution.
    }
    \label{fig:brain-energy}
\end{figure}

\paragraph{\textbf{Damage evolution and internal variable interpretation}}
Figures~\ref{fig:experimental-evaluation}(g–i) show the stress-reduction factor $\raisebox{2pt}{$\chi$}$ inferred by the proposed model as a function of stretch for GM, WM-o, and WM-p, respectively. In all cases, $\raisebox{2pt}{$\chi$}$ decreases gradually from values near unity toward zero as deformation increases, reflecting progressive damage accumulation. Unlike the synthetic examples, where a relatively sharp transition from intact behavior to failure is observed (see Fig. \ref{fig:synthetic_process-level}(b)), the $\raisebox{2pt}{$\chi$}(\lambda)$ curves in this case are smoother and more gradual. This behavior is consistent with the experimental stress–stretch responses, in which the purely elastic regime occupies a small portion of the deformation range and damage accumulates progressively over large stretches before complete failure occurs at extreme stretches ($\lambda \approx 2.25$–$2.8$). Such gradual degradation contrasts with the behavior of many synthetic elastomers, where damage initiates near a critical stretch and progresses rapidly to failure \cite{Hanson2009}. The proposed two-stage framework naturally captures this distinction, discovering the gradual evolution of an internal damage variable directly from experimental data without prescribing its functional form.

\paragraph{\textbf{Prediction of strain energy density saturation and critical failure energy}}
Beyond stress prediction, the proposed model enables estimation of the total strain energy density $\psi$ and, crucially, the critical failure energy $\psi_f$—the energetic limit beyond which the tissue can no longer store mechanical energy. Using the trained models for each dataset, $\psi$ is predicted over an extended stretch range up to $\lambda = 3.5$, exceeding the experimental domain to ensure that complete failure is captured.

Figure~\ref{fig:brain-energy} presents the resulting $\psi$–$\lambda$ curves for the two GM datasets and the two WM substructures. In all cases, $\psi(\lambda)$ increases initially and then asymptotically approaches a constant value, i.e., $\psi_f$, at large stretches, signifying complete failure and energy saturation. This asymptotic behavior arises directly from the constrained GPR learning of the stress-reduction factor $\raisebox{2pt}{$\chi$}(W)$ in Stage~II, which enforces monotonicity, non-negativity, and complete failure.

The corresponding $\psi_f$ values are summarized in Table~\ref{tab:brain_failure_energy}. Notably, despite exhibiting markedly different stress–stretch responses, the two GM datasets converge to nearly identical failure energies (4.22 and 4.28~kPa), yielding an average GM failure energy of 4.25~kPa. By contrast, the average failure energy of WM (WM-o and WM-p) is approximately 2.84~kPa—about 66.8\% of the GM value. This observation suggests a substantially greater resistance of gray matter to mechanical failure compared to white matter, likely stemming from the known structural and compositional differences between these brain substructures \cite{Budday2020}. While additional experimental data are needed to establish robust statistical trends, these results highlight the potential of the proposed framework to uncover meaningful energetic metrics of tissue failure directly from mechanical testing data.
\begin{table}[t]
\centering
\caption{Predicted critical failure energy $\psi_f$ for brain tissue substructures obtained from the proposed two-stage physics-informed GPR-based model. Gray matter (GM) results are reported for two independent specimens, along with their average value. White matter (WM) results correspond to occipital (WM-o) and parietal (WM-p) lobes.}
\label{tab:brain_failure_energy}
\begin{tabular}{lc}
\toprule
\small{\textbf{Tissue type / dataset}} & \small{Critical failure energy, $\psi_f$ (kPa)} \\
\midrule
\small{Gray matter (GM--1)} & \small{4.22} \\
\small{Gray matter (GM--2)} & \small{4.28} \\
\small{\textbf{Gray matter (GM, average)}} & \small{\textbf{4.25}} \\
\midrule
\small{White matter (WM-o)} & \small{2.53} \\
\small{White matter (WM-p)} & \small{3.15} \\
\small{\textbf{White matter (WM, average)}} & \small{\textbf{2.84}} \\
\bottomrule
\end{tabular}
\end{table}

\section{Summary and Conclusions}
\label{sec:summary_and_conclusion}

This work introduced a physics-informed, data-driven constitutive modeling framework for hyperelastic soft materials undergoing progressive damage and failure, with the central objective of reconciling the competing demands of physical interpretability, automation, and predictive accuracy. Motivated by the limitations of classical theory-driven analytical models—whose fixed functional forms restrict flexibility—and purely data-driven black-box models—whose lack of structure undermines generalizability and physical consistency—we developed a two-stage physics-informed GPR-based model that explicitly separates intact elasticity from damage evolution, enforces physics-based constraints, and learns both components directly from data.

The proposed formulation is grounded in the energy-limited hyperelasticity framework, where material failure is characterized by saturation of the strain energy density at a finite critical value. Leveraging the irreducible tensor basis decomposition of stress, the framework expresses the constitutive response in terms of scalar hyperelastic response functions and a stress-reduction factor that governs damage. In Stage I, GPR models are trained to learn the volumetric and isochoric response functions (or only the isochoric response under incompressibility), ensuring exact satisfaction of the principles of determinism and local action, material frame indifference, isotropy, balance of angular momentum, a stress-free reference state, and thermodynamic admissibility through derivation of the intact stress from a scalar strain energy density via the Coleman--Noll procedure. In Stage II, a separate GPR model learns the evolution of the stress-reduction factor as a function of intact strain energy density, with additional physical constraints—non-negativity, monotonic decay, and complete failure at large deformations—implemented through penalty-based constrained optimization to ensure thermodynamic consistency.

Comprehensive validation using synthetic datasets with known ground-truth constitutive behavior demonstrated that the proposed model achieves high in-distribution accuracy under uniaxial tension while substantially outperforming analytical and single-stage GPR-based models in capturing nonlinear softening and failure. Although black-box GPR models achieved the smallest interpolation errors in the training regime, they failed catastrophically when extrapolated to unseen deformation modes. By contrast, the proposed framework exhibited robust out-of-sample generalization to uniaxial compression and simple shear—deformation modes not used during training—successfully reproducing physically expected stress trends, energy saturation, and complete failure. These results highlight a key outcome of the two-stage construction: by operating in a reduced invariant and response-function space and enforcing admissible damage evolution through energetic considerations, the framework generalizes meaningfully beyond the calibration regime, unlike purely data-driven mappings. Notably, this level of accuracy and generalization is achieved using relatively small training datasets consistent with practical experimental availability, underscoring the data efficiency of the proposed framework, an attribute traditionally associated with theory-driven constitutive models.

Beyond predictive accuracy, the framework provides process-level interpretability that is absent in competing ML approaches. The explicit recovery of intact stress, damage evolution, and strain energy density enables a clear physical interpretation of progressive stiffness degradation and failure. In particular, the learned stress-reduction factor exhibits smooth, monotonic decay without spurious oscillations, and the predicted total strain energy density asymptotically approaches a finite saturation value, consistent with energy-limited failure theories. A dedicated investigation of the role of constrained GPR optimization further demonstrated that monotonicity and non-negativity constraints are essential not only for thermodynamic consistency but also for stable and physically meaningful predictions under both training and testing deformation modes.

The practical utility of the framework was demonstrated through its application to experimental uniaxial tension data of brain tissue, a representative soft biological material characterized by noise, high nonlinearity, and limited deformation modes. Under an incompressibility assumption, the model accurately reproduced both elastic and damage-dominated responses across gray and white matter datasets. Importantly, the framework enabled inference of internal damage evolution and prediction of critical failure energy, revealing that gray matter exhibits significantly higher failure energy than white matter—an energetically meaningful distinction that is difficult to extract using conventional constitutive fitting alone. These results underscore the framework’s ability to extract physically relevant material metrics from limited experimental data without prescribing analytical damage laws.

Taken together, this study demonstrates that the proposed two-stage physics-informed GPR-based model combines the best attributes of analytical and data-driven constitutive modeling. Like classical energy-limited hyperelastic formulations, it provides a structured, interpretable, and thermodynamically consistent description of elasticity and damage. At the same time, it retains the automation, flexibility, and high fidelity of modern ML approaches, enabling accurate modeling without expert-driven selection of constitutive forms. As such, the framework represents a significant advancement in the data-driven modeling of failure in soft materials and offers a promising pathway toward automated, physics-consistent constitutive discovery within emerging Material Testing 2.0 and data-centric materials discovery paradigms.

Consistent with classical energy-limited hyperelastic formulations, irreversibility under unloading can be incorporated through a history-dependent switch parameter once a critical energy threshold is reached \cite{Volokh2014}; such an extension is straightforward within the present two-stage framework but is not considered here, as the focus is on monotonic loading to failure.

\paragraph{\textbf{Limitations and future work}}
Despite its advantages, the proposed framework has several limitations that motivate future research. First, although the model is data-driven, it still requires selection of several tunable modeling choices, including the GPR nugget parameter, kernel type, and the cutoff used to isolate the intact response. As demonstrated in the brain tissue application (and further illustrated in Supplementary Section~S1), these choices influence both fitting accuracy and physical interpretability: excessively small nugget values can lead to overfitting and physically inadmissible behavior near the intact stretch cutoff, whereas overly large values may result in excessive smoothing and loss of fidelity in the learned intact response and damage evolution. Developing systematic strategies for automated selection of these modeling choices, sensitivity analysis, and adaptive regularization therefore remains an important direction for future work.

Second, while the two-stage construction and use of GPR provide a degree of interpretability, the physical meaning of GPR parameters—such as kernel structure and characteristic length scales—within the context of solid mechanics is not yet well understood. Establishing clearer connections between these statistical quantities and underlying material behavior would further enhance the interpretability and robustness of the framework.

Finally, classical energy-limited hyperelastic formulations, including those considered here, predict a critical failure energy that is invariant with respect to deformation mode. Our recent work \cite{Chandrashekar2025} has shown that this limitation can be overcome by incorporating deformation-mode dependence through Lode-invariant–based strain energy density formulations. Because the proposed two-stage framework learns the damage evolution law directly from data as a function of a scalar energetic measure, it can be naturally extended to such formulations. More broadly, extending the framework to account for anisotropy, rate dependence, cyclic damage, and thermo-mechanical coupling represents a natural and promising avenue for future research. In addition, building on prior developments within energy-limited hyperelasticity, the present framework can be incorporated into numerical simulations of failure and localization, which is not considered here and will be pursued in future research.





 
\section*{Acknowledgments} \label{Acknowledgments}
This material is based upon work supported by the National Science Foundation under Grant No. 2331294.




\appendix
\setcounter{table}{0}
\renewcommand{\thetable}{A.\arabic{table}}
\renewcommand{\theequation}{A.\arabic{equation}}
\setcounter{equation}{0}

\section{Examples of hyperelastic response functions and stress-reduction factors in energy-limited hyperelasticity}
\label{appendix:stress_derivation}

The generalized stress representation introduced in Section~\ref{sec:Theoretical_Framework} expresses the second Piola–Kirchhoff stress as a weighted sum of irreducible tensor basis components (Eq.~(\ref{eq:12})), where the scalar response functions $\zeta(J)$, $\Gamma_1(\bar{I}_1,\bar{I}_2)$, and $\Gamma_2(\bar{I}_1,\bar{I}_2)$ encode the volumetric and isochoric elastic behavior, and the stress-reduction factor $\raisebox{2pt}{$\chi$}(W)$ governs damage and failure.

In traditional phenomenological constitutive models, both the response functions and the stress-reduction factor assume explicit analytical forms dictated by the chosen constitutive equations for $\psi(W)$, $U(J)$, and $\bar{W}_{iso}(\bar{I}_1,\bar{I}_2)$. Tables~\ref{table:A1} and \ref{table:A2} summarize the resulting response functions for several commonly used volumetric and isochoric hyperelastic models, while Table~\ref{table:A3} lists representative analytical expressions for stress-reduction factors employed in energy-limited damage formulations. Together, these examples illustrate how classical constitutive laws fit naturally within the generalized irreducible tensor basis framework adopted in this work.

Importantly, the proposed physics-informed data-driven framework is not restricted to these fixed functional forms. Instead, it learns the mappings $\zeta(J)$, $\Gamma_1(\bar{I}_1,\bar{I}_2)$, $\Gamma_2(\bar{I}_1,\bar{I}_2)$, and $\raisebox{2pt}{$\chi$}(W)$ directly from data using GPR, thereby overcoming the limited flexibility of explicit analytical expressions while preserving physical structure, thermodynamic consistency, and interpretability.

\begin{table}[ht]
    \centering 
    \caption{Response function $\zeta(J)$ for commonly employed volumetric energy density functions ($U(J)$): the Simo--Miehe model \cite{Simo1992}, the volumetric neo-Hookean model \cite{DeRooij2016}, and the volumetric Ogden model \cite{ogden1997}.}
    \label{table:A1}
    \begin{tabular}{lcc}
    \toprule
    \small{\textbf{Model}}               & \small{$U$}             & \small{$\zeta$}                                      \\
    \midrule
    \small{Simo--Miehe model}             & \small{$\frac{\kappa}{2} \left( \frac{J^2 - 1}{2} - \ln{J} \right)$}               & \small{$\frac{\kappa}{2} \left( J^2 - 1 \right)$}                 \\
    \small{Volumetric neo-Hookean model} & \small{$\frac{\kappa}{2} \left(J - 1\right)^2$}                                    & \small{$\kappa J \left( J - 1 \right)$}                             \\
    \small{Volumetric Ogden model}        & \small{$\frac{\kappa}{\beta^2} \left(\frac{1}{J^\beta} - 1 + \beta \ln{J}\right)$} & \small{$\frac{\kappa}{\beta} \left( 1 - \frac{1}{J^\beta} \right)$} \\
    \bottomrule
    \end{tabular}
\end{table}
\begin{table}[ht]
    \centering
    \caption{Response functions $\Gamma_1(\bar{I}_1,\bar{I}_2)$ and $\Gamma_2(\bar{I}_1,\bar{I}_2)$ for commonly employed isochoric hyperelastic strain energy density functions ($\bar{W}_{iso}(\bar{I}_1,\bar{I}_2)$): the neo-Hookean model \cite{Treloar1943,Rivlin1948a}, the Mooney--Rivlin model \cite{Mooney1940, Rivlin1948b}, the Yeoh model \cite{Yeoh1993}, the Gent model \cite{Gent1996}, and the Gent--Gent model \cite{Pucci2002}.}
    \label{table:A2}
        \renewcommand{\arraystretch}{1.15}
    \setlength{\tabcolsep}{3pt}
    \begin{tabular}{lccc}
    \toprule
    \small{\textbf{Model}}          & \small{$\bar{W}_{iso}$}                                                                                                           & \small{$\Gamma_1$}                                                             & \small{$\Gamma_2$}                          \\
    \midrule
    \small{Neo-Hookean model}       & \small{$A_{10}\left(\Bar{I}_1 - 3\right)$}                                                                                                 & \small{$2A_{10}$}                                                                           & \small{$0$}                                              \\
    \small{Mooney--Rivlin model}     & \small{$A_{10}\left(\Bar{I}_1 - 3\right) + A_{01}\left(\Bar{I}_2 - 3\right)$}                                                              & \small{$2 \left( A_{10} + \Bar{I}_1 A_{01}\right)$}                                         & \small{$- 2 A_{01}$}                                     \\
    \small{Yeoh model}               & \small{$C_1 \left(\Bar{I}_1 - 3\right) + C_2 \left(\Bar{I}_1 - 3\right)^2$}                                                                & \small{$2 C_1 +  4 C_2 \left(\Bar{I}_1 - 3\right)$}                                         & \small{$0$}                                              \\
    \small{Gent model}               & \small{$-\frac{\mu J_m}{2}\ln{\left(1 - \frac{\Bar{I}_1 - 3}{J_m}\right)}$}                                                                & \small{$\frac{\mu}{1 - \frac{\Bar{I}_1 - 3}{J_m}}$}                                         & \small{$0$} \\
    \small{Gent--Gent model}         & \small{$-\frac{\mu J_m}{2}\ln{\left(1 - \frac{\Bar{I}_1 - 3}{J_m}\right)} + \frac{3C_2}{2}\ln{\left(\frac{\Bar{I}_2}{3}\right)}$}          & \small{$\frac{\mu}{1 - \frac{\Bar{I}_1 - 3}{J_m}} + 3 C_2 \frac{\Bar{I}_1}{\Bar{I}_2}$}     & \small{$-3\frac{C_2}{\Bar{I}_2}$} \\
    \bottomrule
    \end{tabular}
\end{table}
\begin{table}[ht]
    \centering 
    \caption{Stress-reduction factor $\raisebox{2pt}{$\chi$}(W)$ for available energy-limited hyperelastic models ($\psi(W)$): the Volokh reduced damage model \cite{Volokh2007}, the Volokh universal damage model \cite{Volokh2010}, and the Chandrashekar--Upadhyay model \cite{Chandrashekar2025}.}
    \label{table:A3}
    \renewcommand{\arraystretch}{1.15}
    \setlength{\tabcolsep}{3pt}
    \begin{threeparttable}
    
    \begin{tabular}{p{1.4in} >{\raggedright\arraybackslash}m{3in} >{\raggedright\arraybackslash}m{1.35in}}
    \toprule
    \small{\textbf{Model}}               & \small{$\psi$}             & \small{$\raisebox{2pt}{$\chi$}$}                                      \\
    \midrule
    \small{Volokh reduced damage model}             & \small{$\Phi - \Phi\mathrm{exp}\left(-\frac{W}{\Phi}\right)$}               & \small{$\exp\left(-\frac{W}{\Phi}\right)$} \\
    \small{Volokh universal damage model} & \small{$\frac{\Phi}{m}\left\{ \mathit{\Gamma^*}\left( \frac{1}{m},0 \right) - \mathit{\Gamma^*}\left( \frac{1}{m}, \left(\frac{W}{\Phi}\right)^m \right)\right\}$}                                    & \small{$\exp\left(-\left(\frac{W}{\Phi}\right)^m\right)$}                             \\
    \small{Chandrashekar--Upadhyay model\tnote{$\dagger$}}        & \small{$
\frac{\Phi^{-}}{m^{-}}\,\mathit{\Gamma^*}\!\left(\frac{1}{m^{-}},0\right)
+ \beta\!\left[
\frac{\Phi^{+}}{m^{+}}\,\mathit{\Gamma^*}\!\left(\frac{1}{m^{+}},0\right)
- \frac{\Phi^{-}}{m^{-}}\,\mathit{\Gamma^*}\!\left(\frac{1}{m^{-}},0\right)
\right]
-\frac{\Phi^{-}}{m^{-}}\,\mathit{\Gamma^*}\!\left(\frac{1}{m^{-}},\left(\frac{W}{\Phi^{-}}\right)^{m^{-}}\right)
-\beta\!\left[
\frac{\Phi^{+}}{m^{+}}\,\mathit{\Gamma^*}\!\left(\frac{1}{m^{+}},\left(\frac{W}{\Phi^{+}}\right)^{m^{+}}\right)
- \frac{\Phi^{-}}{m^{-}}\,\mathit{\Gamma^*}\!\left(\frac{1}{m^{-}},\left(\frac{W}{\Phi^{-}}\right)^{m^{-}}\right)
\right]$}               & \small{$\beta \mathrm{exp}\left(-\left(\frac{W}{\Phi^+}\right)^{m^+}\right) + (1-\beta) \mathrm{exp}\left(-\left(\frac{W}{\Phi^-}\right)^{m^-}\right)$} \\
    \bottomrule
    \end{tabular}

    \begin{tablenotes}
        \scriptsize{\item[$\dagger$] In the Chandrashekar--Upadhyay model, $\beta$ is a Lode-angle-dependent weighting function introduced to capture deformation-mode dependence (e.g., tension-, compression-, and shear-dominated loading) of damage and failure.}
    \end{tablenotes}

    \end{threeparttable}
\end{table}

\section{Tangent stiffness tensor for energy-limited hyperelasticity}
\label{appendix:stiffness}
In this appendix, we derive the fourth-order material tangent stiffness tensor (also referred to as the elasticity tensor in the material description), denoted by \( \mathbb{C} \), for the general class of energy-limited hyperelastic constitutive models considered in this work. The resulting expression is independent of the specific functional forms adopted for the hyperelastic response functions and stress-reduction factor, and is therefore applicable to both analytical and data-driven formulations within the energy-limited hyperelasticity framework.

Within the energy-limited hyperelasticity framework, the second Piola--Kirchhoff stress tensor admits the generalized representation (same as Eq.~(\ref{eq:12})):
\begin{equation}
\mathbf{S} = \chi(W) \left( \zeta(J) \mathbb{G}_1 + J^{-2/3} \Gamma_1(\bar{I}_1,\bar{I}_2) \mathbb{G}_2 + J^{-2/3} \Gamma_2(\bar{I}_1,\bar{I}_2) \mathbb{G}_3 \right),
\end{equation}
where \( \mathbb{G}_1 = \mathbf{C}^{-1} \), \( \mathbb{G}_2 = \mathrm{Dev}(\mathbf{I}) \), and \( \mathbb{G}_3 = \mathrm{Dev}(\bar{\mathbf{C}}) \), with the deviatoric operator defined as:
\begin{equation}
\mathrm{Dev}(\mathbf{A}) = \mathbf{A} - \frac{1}{3}(\mathbf{A} : \mathbf{C})\mathbf{C}^{-1}.
\end{equation}

The fourth-order material tangent stiffness tensor is defined as \cite{Holzapfel2000}
\begin{equation}
\label{eq:A3}
\mathbb{C} = 2 \frac{\partial \mathbf{S}}{\partial \mathbf{C}}.
\end{equation}
Substituting the generalized stress representation into this definition yields
\begin{equation}
\label{eq:A4}
\mathbb{C} = 2 \frac{\partial}{\partial \mathbf{C}} \left( \chi \left( \zeta \mathbb{G}_1 + J^{-2/3} \Gamma_1 \mathbb{G}_2 + J^{-2/3} \Gamma_2 \mathbb{G}_3 \right) \right).
\end{equation}

For notational simplicity, the explicit functional dependence of the scalar response functions and the stress-reduction factor on their respective arguments (i.e., invariants and strain energy density) is omitted in the remainder of this appendix. Further, we introduce the intact (undamaged) stress tensor, defined as the derivative of the intact strain energy density (same as Eq. (\ref{eq:20})),
\begin{equation}
\label{eq:A5}
\mathbf{S}_{\mathrm{intact}} := 2\frac{\partial W}{\partial \mathbf{C}}
= \zeta\mathbb{G}_1
+ J^{-2/3}\Gamma_1\mathbb{G}_2
+ J^{-2/3}\Gamma_2\mathbb{G}_3.
\end{equation}

Applying the chain rule to Eq. (\ref{eq:A4}) and using Eq. (\ref{eq:A5}), the tangent stiffness tensor can be expanded as
\begin{equation}
\label{eq:A6}
\begin{aligned} 
\mathbb{C} = & \frac{d\chi}{dW} \mathbf{S}_{\mathrm{intact}} \otimes \mathbf{S}_{\mathrm{intact}} \\
& + 2 \chi \Bigg[ \mathbb{G}_1 \otimes \frac{d\zeta}{dJ} \frac{\partial J}{\partial \mathbf{C}} + \zeta \frac{\partial \mathbb{G}_1}{\partial \mathbf{C}} + \Gamma_1 \mathbb{G}_2 \otimes\left( -\frac{2}{3} J^{-5/3} \frac{\partial J}{\partial \mathbf{C}} \right) \\ 
& \qquad \quad + J^{-2/3} \mathbb{G}_2 \otimes \left( \frac{\partial \Gamma_1}{\partial \bar{I}_1} \frac{\partial \bar{I}_1}{\partial \mathbf{C}} + \frac{\partial \Gamma_1}{\partial \bar{I}_2} \frac{\partial \bar{I}_2}{\partial \mathbf{C}} \right) + J^{-2/3} \Gamma_1 \frac{\partial \mathbb{G}_2}{\partial \mathbf{C}} \\
& \qquad \quad + \Gamma_2 \mathbb{G}_3 \otimes\left( -\frac{2}{3} J^{-5/3} \frac{\partial J}{\partial \mathbf{C}} \right) + J^{-2/3} \mathbb{G}_3 \otimes \left( \frac{\partial \Gamma_2}{\partial \bar{I}_1} \frac{\partial \bar{I}_1}{\partial \mathbf{C}} + \frac{\partial \Gamma_2}{\partial \bar{I}_2} \frac{\partial \bar{I}_2}{\partial \mathbf{C}} \right) \\
& \qquad \quad + J^{-2/3} \Gamma_2 \frac{\partial \mathbb{G}_3}{\partial \mathbf{C}} \Bigg],
\end{aligned}
\end{equation}
where the standard dyadic product $\otimes$ is defined as $(\mathbf{A} \otimes \mathbf{B})_{ijkl} = A_{ij} B_{kl}.$ The first term in the above expression arises from the dependence of the stress-reduction factor $\raisebox{2pt}{$\chi$}(W)$ on the intact strain energy density $W$. Since $\mathbf{S}_{\mathrm{intact}} = 2\,\partial W / \partial \mathbf{C}$, this contribution represents the energetic coupling between damage evolution and elastic deformation.
    
The explicit expressions for the derivative terms in Eq. (\ref{eq:A6}) needed to compute $\mathbb{C}$ are provided below:
\begin{itemize}
\item Derivative of \( J \) with respect to \( \mathbf{C} \):
\begin{equation}
\frac{\partial J}{\partial \mathbf{C}} = \frac{1}{2} J \mathbf{C}^{-1}.
\end{equation}

\item Derivative of \( \mathbb{G}_1 \) with respect to \( \mathbf{C} \):
\begin{equation}
\frac{\partial \mathbb{G}_1}{\partial \mathbf{C}} = -\mathbf{C}^{-1} \odot
 \mathbf{C}^{-1},
\end{equation}
where the symbol $\odot$ denotes tensor product following the rule $\left(\mathbf{A}\odot\mathbf{B}\right)_{ijkl}=\frac{1}{2}\left(\mathbf{A}_{ik}\mathbf{B}_{jl} + \mathbf{A}_{il}\mathbf{B}_{jk}\right)$ \cite{Holzapfel2000}.

\item Derivative of \( \bar{I}_1 \) and \( \bar{I}_2 \) with respect to \( \mathbf{C} \):
\begin{equation}
\frac{\partial \bar{I}_1}{\partial \mathbf{C}} = J^{-2/3} \left( \mathbf{I} - \frac{1}{3}(\mathrm{tr}\mathbf{C}) \mathbf{C}^{-1} \right),
\end{equation}
\begin{equation}
\frac{\partial \bar{I}_2}{\partial \mathbf{C}} = J^{-4/3} \left[ (\mathrm{tr}\mathbf{C})\mathbf{I} - \mathbf{C} - \frac{\mathbf{C}^{-1}}{3} \left( (\mathrm{tr}\mathbf{C})^2 - \mathrm{tr}(\mathbf{C}^2)\right) \right].
\end{equation}
    
\item Derivative of \( \mathbb{G}_2 \) with respect to \( \mathbf{C} \):
\begin{equation}
\frac{\partial \mathbb{G}_2}{\partial \mathbf{C}} = -\frac{1}{3} \left[ \mathbf{C}^{-1} \otimes \mathbf{I} - \mathrm{tr}\mathbf{C} \left( \mathbf{C}^{-1} \odot \mathbf{C}^{-1} \right)\right].
\end{equation}

\item Derivative of \( \mathbb{G}_3 \):
\begin{equation}
\frac{\partial \mathbb{G}_3}{\partial \mathbf{C}} = J^{-2/3} \left[ \mathbb{I} - \frac{2}{3} \mathbf{C}^{-1} \otimes \mathbf{C} + \frac{1}{3} \mathrm{tr}(\mathbf{C}^2) \left( \mathbf{C}^{-1} \odot \mathbf{C}^{-1} \right) - \frac{1}{3} \mathbf{C} \otimes \mathbf{C}^{-1} + \frac{1}{9} \mathrm{tr}(\mathbf{C}^2) \left( \mathbf{C}^{-1} \otimes \mathbf{C}^{-1} \right) \right],
\end{equation}
where $\mathbb{I}$ denotes the fourth-order unit tensor ($\mathbb{I}_{ijkl} = \delta_{ik} \delta_{jl}$).

\end{itemize}

This completes the derivation of the material tangent stiffness tensor $\mathbb{C}$ for energy-limited hyperelastic constitutive models expressed in terms of irreducible tensor bases, strain invariants, scalar response functions, and a stress-reduction factor. The corresponding spatial (Eulerian) tangent stiffness tensor $\mathbbm{c}$—also referred to as the elasticity tensor in the spatial description—can be obtained through a standard push-forward operation as \cite{Holzapfel2000}
\begin{equation}
\label{eq:A13}
\mathbbm{c}_{ijkl} = J^{-1} F_{ip} F_{jq} F_{kr} F_{ls} \, \mathbb{C}_{pqrs},
\end{equation}
where $\mathbf{F}$ is the deformation gradient.

Because the above derivation of tangent stiffness tensors does not rely on any specific functional forms, the resulting tangent operators apply equally to analytical and data-driven formulations within the energy-limited hyperelasticity framework. This provides the constitutive linearization required for numerical implementation (e.g., finite element analysis) of the proposed two-stage physics-informed GPR-based model.

 \bibliographystyle{elsarticle-num} 
 \bibliography{cas-refs}





\end{document}


\maketitle

\begin{center}
\large \textbf{Supplementary Material}
\end{center}
\vspace{1em}

\section*{S1\quad Effect of the nugget parameter in the Stage I isochoric GPR model}
\setcounter{figure}{0}
\renewcommand{\thefigure}{S\arabic{figure}}

This section examines the sensitivity of the proposed framework to the choice of the nugget parameter $\alpha$ used in the Stage~I isochoric GPR model $\mathcal{M}_{iso}$. All other modeling choices, including kernel type, training data, intact cutoff stretch, and Stage~II constrained damage learning, are held fixed.

\begin{figure}[t]
    \centering
    \includegraphics[width=5in]{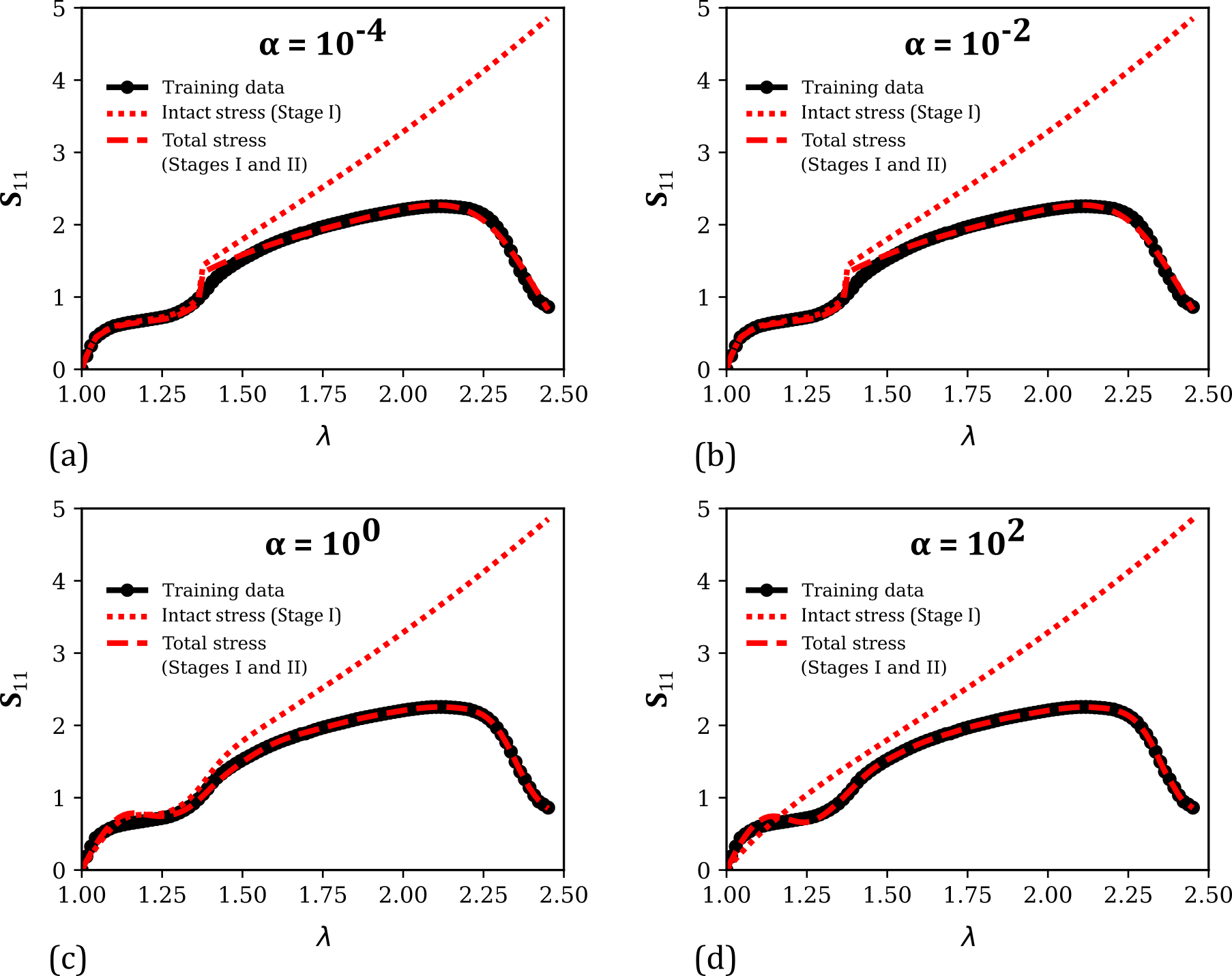}
    \caption{%
    Sensitivity of intact and total axial stress--stretch predictions to the nugget parameter $\alpha$ in the Stage~I isochoric GPR model $\mathcal{M}_{iso}$. Results are shown for $\alpha = 10^{-4}$, $10^{-2}$, $10^{0}$, and $10^{2}$. Very small nugget values lead to overfitting and a spurious stress discontinuity at the intact cutoff stretch, which propagates into the total stress response. Intermediate values eliminate this discontinuity and yield smooth, physically meaningful predictions, while excessively large nugget values result in over-smoothing and loss of interpretability.
    }
    \label{fig:supp_nugget_sensitivity}
\end{figure}

Figure~\ref{fig:supp_nugget_sensitivity} compares intact and total axial stress--stretch predictions obtained using four representative nugget values, $\alpha = 10^{-4}$, $10^{-2}$, $10^{0}$, and $10^{2}$, for a representative brain tissue gray matter dataset. For very small nugget values ($\alpha = 10^{-4}$ and $10^{-2}$), the intact response interpolates the training data almost exactly; however, this near-interpolatory behavior introduces a spurious stress discontinuity at the intact cutoff stretch. This discontinuity propagates into the total stress prediction, is physically inadmissible, leads to elevated global errors, and may cause numerical issues in simulations due to the associated stiffness tensor.

As the nugget parameter is increased, these overfitting-induced artifacts are progressively alleviated. For $\alpha = 10^{0}$, the intact and total stress responses become smooth across the intact cutoff, and the spurious stress jump vanishes. Although a mild overprediction near the transition from elastic to damage-dominated behavior is observed—consistent with the discussion in the main manuscript—the overall response remains physically reasonable and numerically well behaved. Based on systematic evaluation across all brain tissue datasets considered in this work, an intermediate nugget value $\alpha = 0.5$ was found to provide the best balance between smoothness, fidelity, and physical interpretability.

For excessively large nugget values ($\alpha = 10^{2}$), the intact response becomes over-smoothed and deviates noticeably from the total response even within the nominally intact regime. As a result, the inferred stress-reduction factor $\raisebox{2pt}{$\chi$}$ significantly deviates from unity at small strains, diminishing the physical interpretability of the elastic--damage decomposition.

Overall, this sensitivity study highlights that appropriate nugget selection in Stage~I is essential not only for fitting accuracy, but also for avoiding unphysical stress discontinuities, ensuring numerical robustness, and preserving meaningful damage evolution in the proposed two-stage framework.
